\documentclass[a4paper,fleqn,usenatbib]{mnras}

\usepackage[T1]{fontenc}

\usepackage[english]{babel}
\usepackage{graphicx}
\usepackage{fleqn}
\usepackage{amssymb}
\usepackage{amsmath}
\usepackage{booktabs}
\usepackage{hyperref}
\usepackage{threeparttable}
\usepackage{txfonts} 

\usepackage{color, colortbl}

\renewcommand{\S}{Section}
\newcommand{\F}{Fig.}

\newcommand{\ve}[1]{\boldsymbol{#1}}

\title[HJs in triples]{On the formation of hot and warm Jupiters via secular high-eccentricity migration in stellar triples}

\author[Hamers]{Adrian S. Hamers$^{1}$\thanks{E-mail: hamers@ias.edu}  \\
$^{1}$Institute for Advanced Study, School of Natural Sciences, Einstein Drive, Princeton, NJ 08540, USA}
\date{Accepted 2017 January 5. Received 2016 December 5; in original form 2016 October 21}

\pubyear{2017}

\begin{document}
\label{firstpage}
\pagerange{\pageref{firstpage}--\pageref{lastpage}}
\maketitle

\begin{abstract} 
Hot Jupiters (HJs) are Jupiter-like planets orbiting their host star in tight orbits of a few days. They are commonly believed not to have formed {\it in situ}, requiring inwards migration towards the host star. One of the proposed migration scenarios is secular high-eccentricity or high-$e$ migration, in which the orbit of the planet is perturbed to high eccentricity by secular processes, triggering strong tidal evolution and orbital migration. Previous theoretical studies have considered secular excitation in stellar binaries. Recently, a number of HJs have been observed in stellar triple systems. In the latter, the secular dynamics are much richer compared to stellar binaries, and HJs could potentially be formed more efficiently. Here, we investigate this possibility by modeling the secular dynamical and tidal evolution of planets in two hierarchical configurations in stellar triple systems. We find that the HJ formation efficiency is higher compared to stellar binaries, but only by at most a few tens of per cent. The orbital properties of the HJs formed in the simulations are very similar to HJs formed in stellar binaries, and similarly to studies of the latter we find no significant number of warm Jupiters. HJs are only formed in our simulations for triples with specific orbital configurations, and our constraints are approximately consistent with current observations. In future, this allows to rule out high-$e$ migration in stellar triples if a HJ is detected in a triple grossly violating these constraints. 
\end{abstract}


\section{Introduction}
\label{sect:introduction}
Hot Jupiters (HJs) are Jupiter-like planets orbiting their host star on tight orbits, downward of 10 d. Despite the first detection of a HJ over two decades ago \citep{1995Natur.378..355M}, it is still unclear how these planets could have formed. It is commonly believed that planet formation in the protoplanetary disk phase is not efficient so close to the star (e.g. \citealt{1996Natur.380..606L}), although some recent studies have considered the possibility of {\it in situ} formation through core accretion \citep{2014ApJ...797...95L,2016ApJ...817L..17B,2016ApJ...829..114B}. Discounting the latter possibility, then the proto-HJ must have formed in regions further away from the star, i.e. beyond the ice line of one to several AU, and subsequently migrated inwards. 

Two main migration scenarios have been considered: (1) disk migration e.g. \citealt{1980ApJ...241..425G,1986ApJ...309..846L,2000Icar..143....2B,2002ApJ...565.1257T}), and (2) migration induced by tidal dissipation in the HJ, requiring high orbital eccentricity, also known as `high-$e$' migration. Several mechanisms have been proposed to drive the high eccentricity required for high-$e$ migration, including
\begin{enumerate}
\item close encounters between planets \citep{1996Sci...274..954R,2008ApJ...686..580C,2008ApJ...686..621F,2008ApJ...686..603J,2008ApJ...678..498N,2012ApJ...751..119B};
\item secular Lidov-Kozai (LK) oscillations \citep{1962P&SS....9..719L,1962AJ.....67..591K} induced by a distant companion star or an additional (massive) planet on an inclined orbit \citep{2003ApJ...589..605W,2007ApJ...669.1298F,2012ApJ...754L..36N,2015ApJ...799...27P,2016MNRAS.456.3671A,2016ApJ...829..132P};
\item secular excitation induced by a close and coplanar, but eccentric planetary companion \citep{2015ApJ...805...75P,2016ApJ...820...55X}, and
\item secular chaos in multiplanet systems with at least three planets in mildly inclined and eccentric orbits \citep{2011ApJ...735..109W,2011ApJ...739...31L,2014PNAS..11112610L,2016MNRAS.tmp.1471H}. 
\end{enumerate}

\begin{figure*}
\centering
\includegraphics[scale = 0.75, trim = 0mm 0mm 0mm 0mm]{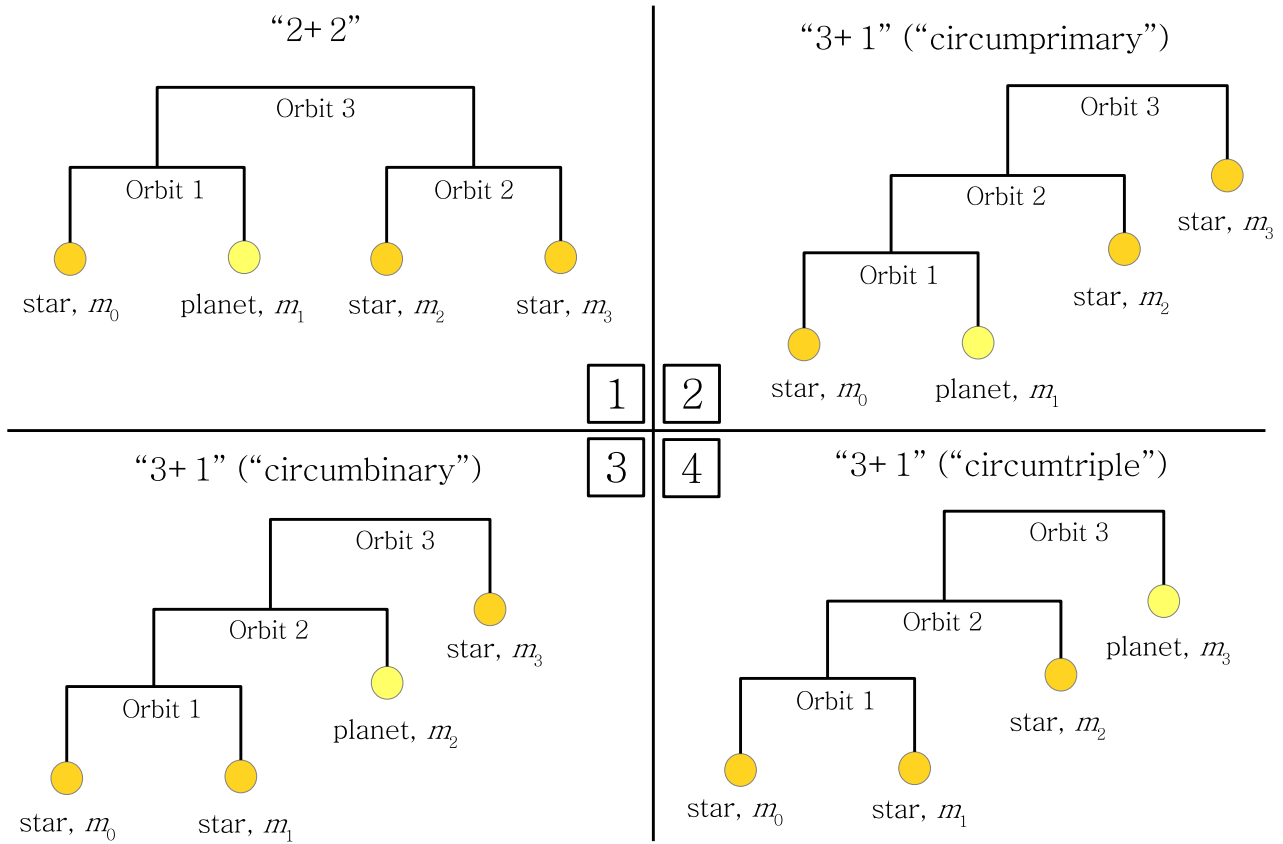}
\caption{\small Schematic representations of hierarchical orbits of planets in stellar triple systems, using mobile diagrams \citep{1968QJRAS...9..388E}. Note that these diagrams only depict the hierarchy of the system, and not the relative sizes and orientations of the orbits. For hierarchical four-body systems, there are generally two distinct hierarchical configurations (`2+2' and `3+1'). However, taking into account that the stars are distinct from the planet, there are in this case four physically distinct configurations shown in each of the panels. In this work, only the configurations in panels 1 and 2 are considered in the context of HJ formation through secular high-$e$ migration.  }
\label{fig:configurations.eps}
\end{figure*}

The orbital periods of HJs peak around $\sim 3-5 \, \mathrm{d}$. In the case of disk migration, migration in principle proceeds until the planet is engulfed by the star, and truncation of the disk needs to be invoked in order to explain the observed orbital period distribution. In contrast, high-$e$ migration induced by secular processes naturally predicts a `stalling' orbital period of  $\sim 3-5 \, \mathrm{d}$. Also, for disk migration the obliquity (the angle between the stellar spin and orbit of the planet) is expected to be (close to) zero, unless the stellar spin axis was initially misaligned with respect to the protoplanetary disk (e.g. \citealt{2010MNRAS.401.1505B,2011MNRAS.412.2799F,2012Natur.491..418B,2014MNRAS.440.3532L}). In contrast, large obliquities, even retrograde, have been observed (e.g. \citealt{2011AJ....141...63W,2015ApJ...801....3M}), and these are more naturally explained in the case of secular high-$e$ migration, in which the orbit of the planet is continuously changing its orientation with respect to the star before the evolution is dominated by tidal dissipation.

However, secular high-$e$ migration generally faces the problem that the theoretical occurrence rates are about an order of magnitude too low compared to observations. Also, a high efficiency of tidal dissipation in the HJ is required, and the predicted orbital periods are on the short end of the observations unless tidal dissipation is very efficient (e.g. \citealt{2015ApJ...799...27P,2016MNRAS.456.3671A}).

Recently, HJs have been found orbiting stars in stellar triple systems. The three HJs are WASP-12b \citep{2009ApJ...693.1920H,2013MNRAS.428..182B,2014ApJ...788....2B}, HAT-P-8b \citep{2009ApJ...704.1107L,2013MNRAS.428..182B,2014ApJ...788....2B}, and KELT-4Ab \citep{2016AJ....151...45E}. In these systems, the HJs orbit the tertiary star in a stellar triple system, i.e. the HJ host star is orbited by a more distant pair of stars. From a dynamical point of view, such a system is a hierarchical four-body system in the `2+2'  or `binary-binary' configuration, where one of the binaries is composed of a single star + planet, and the other binary is composed of the two other stars. The binarity of the distant pair can affect the orbital evolution of the planet around its host star. In particular, \citet{2013MNRAS.435..943P} showed that in these systems the parameter space for exciting high eccentricities triggered by flips of the orbital planes is larger compared to the equivalent triple systems, potentially enhancing the formation rate of HJs compared to the case of a single stellar companion. 

The observations of HJs in stellar triples mentioned above constitute only one example of the configurations in which planets can orbit stars in stellar triple systems. For hierarchical four-body systems, there exist two classes of long-term stable configurations: the `2+2' and the `3+1' configurations\footnote{Here, we do not consider the possibility of non-hierarchical orbits such as trojan orbits.}. In the `2+2' or `binary-binary' configuration (cf. panel 1 of \F\,\ref{fig:configurations.eps}), two binary pairs are bound in a wider orbit. In the `3+1' or `triple-single' configuration (cf. panels 2-4 of \F\,\ref{fig:configurations.eps}), a hierarchical triple, composed of an inner and outer orbit, is orbited by a more distant fourth body. 

In the case of `2+2' systems, the planet orbits one of the stars, and only one physically unique configuration is possible (cf. panel 1 of \F\,\ref{fig:configurations.eps}). For `3+1' systems, three physically distinct configurations exist. The planet can orbit
\begin{enumerate}
\item one of the stars in the innermost binary (i.e. an `S-type' or `circumprimary' orbit; cf. panel 2 of \F\,\ref{fig:configurations.eps});
\item the (center of mass of the) innermost binary (i.e. a `P-type' or `circumbinary' orbit; cf. panel 3 of \F\,\ref{fig:configurations.eps});
\item the (center of mass of the) triple system (i.e. a `circumtriple' orbit; cf. panel 4 of \F\,\ref{fig:configurations.eps}).
\end{enumerate}
Note that the nomenclature `S-type' and `P-type', introduced by \citet{1982OAWMN.191..423D}, strictly only applies to binary star systems. 

In case (i), the planet is orbiting a single star; it can approach this star at a short distance (i.e. a fraction of an AU) while maintaining dynamical stability, i.e. not destroying the hierarchy of the system. However, in cases (ii) and (iii), the orbit of the planet is very likely unstable if the planet were to approach one of the stars at such a short distance because of the perturbations by the other stars. Therefore, for `3+1' systems, likely only case (i) is relevant for HJ formation through secular high-$e$ migration because the latter requires repeated close passages to a star. Nevertheless, it might be possible to form HJs through tidal capture of dynamically unstable planets by stars, triggered e.g. by secular evolution. We do not consider the latter possibility here. In case (i), similarly to `2+2' four-body systems, coupled LK oscillations  between the orbits can result in extremely high eccentricities and secular chaotic motion \citep{2015MNRAS.449.4221H}. 

The above considerations suggest a higher formation efficiency of HJs in stellar triples compared to stellar binaries\footnote{Not taking into account the relative frequencies of binaries and triples.}, and this motivates study of the efficiency of HJ formation through secular high-$e$ migration in stellar triples. In this paper, we investigate this through numerical simulations, focussing on the `2+2' and `3+1' (i) configurations (i.e. panels 1 and 2 of  \F\,\ref{fig:configurations.eps}). We pay particular emphasis on comparing the efficiency of HJ production to the case of a stellar binary (i.e. panels 1 or 2 of \F\,\ref{fig:configurations.eps} with $m_3$ removed). 

Furthermore, we consider the conditions necessary for HJ formation through high-$e$ migration in triples, and show that these conditions exclude the formation of HJs through high-$e$ migration in triples with certain configurations. Although not presently the case, a future detection of a HJ violating these conditions would be strong indication for another formation mechanism at work such as disk migration.

The structure of this paper is as follows. In \S\,\ref{sect:methods}, we describe the secular method and other assumptions. We give an example of the rich secular dynamics of planets in stellar triples in \S\,\ref{sect:example}. In \S\,\ref{sect:pop_syn}, we present the results from the population synthesis study. We discuss our results in \S\,\ref{sect:discussion}, and conclude in \S\,\ref{sect:conclusions}.

\section{Methods and assumptions}
\label{sect:methods}

\begin{table}
\begin{threeparttable}
\begin{tabular}{lp{3.8cm}c}
\toprule
Symbol & Description & (Range of) value(s) \\
\midrule
$m_0$                   & Stellar mass                  & $1\,\mathrm{M}_\odot$                                 \\
$m_1$                    & Planetary mass             & 1 $M_\mathrm{J}$ \\
$m_2$                    & Stellar mass             & $q_2 m_0$ ${}^\mathrm{a}$\\
$m_3$                    & Stellar mass             & $q_3 m_0$ ${}^\mathrm{a}$ \\
$R_0$                   & Stellar radius                & $1\,\mathrm{R}_\odot$    \\
$R_1$                      & Planetary radius         & $1\,R_\mathrm{J}$  \\
$R_2$                      & Stellar radius         & $(m_2/\mathrm{M}_\odot)^{0.8} \, \mathrm{R}_\odot$  ${}^\mathrm{b}$ \\
$R_3$                      & Stellar radius         & $(m_3/\mathrm{M}_\odot)^{0.8} \, \mathrm{R}_\odot$  ${}^\mathrm{b}$ \\
$\eta$                      & Tidal disruption factor   & 2.7 ${}^\mathrm{c}$ \\
$t_\mathrm{V,\star}$        & Stellar viscous time-scale                    & $5 \, \mathrm{yr}$ \\
$t_\mathrm{V,1}$            & Planetary viscous time-scale     & $0.0137,0.137,1.37\,\mathrm{yr}$ \\
$k_\mathrm{AM,\star}$       & Stellar apsidal motion constant & 0.014 \\
$k_\mathrm{AM,1}$           & Planetary apsidal motion constant                    & 0.25 \\
$r_\mathrm{g,\star}$        & Stellar gyration radius                                                & 0.08 \\
$r_\mathrm{g,1}$            & Planetary gyration radius                                               & 0.25 \\
$P_\mathrm{s,\star}$        & Stellar spin period                                                              & $10 \, \mathrm{d}$ \\
$P_\mathrm{s,1}$            & Planetary spin period                                                               & $10 \, \mathrm{hr}$ \\
$\theta_0$              & Stellar obliquity (stellar spin-planetary orbit angle)                                    & $0^\circ$ \\
$a_1$                       & Planetary orbital semimajor axis & 1-4 AU \\
$P_2$ 			& Period of orbit 2 & 1-$10^{10}$ d ${}^\mathrm{d}$\\
$P_3$ 			& Period of orbit 3 & 1-$10^{10}$ d ${}^\mathrm{d}$ \\
$e_1$                       & Planetary orbital eccentricity & 0.01 \\
$e_i$                       & Orbit $i$ eccentricity ($i \neq 1$) & 0.01 - 0.9 ${}^\mathrm{e}$ \\
$i_1$ 			& Planetary orbital inclination & 0${}^\circ$ \\
$i_i$ 			& Orbit $i$ inclination ($i \neq 1$) & 0-180${}^\circ$ \\
$\omega_i$ & Orbit $i$ Argument of pericenter              & 0-360${}^\circ$                                          \\
$\Omega_i$ & Orbit $i$ Longitude of ascending node         & 0-360${}^\circ$                                     \\
\bottomrule
\end{tabular}
\begin{tablenotes}
            \item[a] Mass ratio $q_i$ sampled from a flat distribution between 0.1 and 1.0.
            \item[b] \citet{1994sse..book.....K}.
            \item[c] \citet{2011ApJ...732...74G}; cf. equation~\ref{eq:r_t}).
            \item[d] Lognormal distribution \citep{2010ApJS..190....1R}.
            \item [e] Rayleigh distribution with an rms width of 0.33 \citep{2010ApJS..190....1R}.
\end{tablenotes}
\caption{Description of quantities relevant for \S\,\ref{sect:pop_syn} (see also \F\,\ref{fig:configurations.eps}) and their assumed (ranges of) value(s). The stellar quantities with the `$\star$' subscript apply to all three stars, i.e. $i\in \{0,2,3\}$. The orbital elements are defined with respect to an arbitrary reference frame. Note that the ranges of orbital parameters are the ranges before applying dynamical stability constraints. }
\label{table:IC}
\end{threeparttable}
\end{table}

\subsection{Configurations and notation}
In panels 1 and 2 of \F\,\ref{fig:configurations.eps}, the hierarchical configurations of planets in triples considered in this work are depicted schematically. In addition to these two configurations, we included the case of a stellar binary, i.e. panels 1 or 2 of \F\,\ref{fig:configurations.eps} with $m_3$ removed. Invariably, we use $m_0$ to denote the mass of the star that the planet is orbiting, and $m_1$ to denote the planetary mass. The other two stellar masses are denoted with $m_2$ and $m_3$. Depending on the configuration, there are two to three orbits in the system, labeled 1 to 3. For example, for `2+2' systems, the semimajor axis of the orbit of the planet around its star is denoted with $a_1$, the semimajor axis of the orbit of stars $m_2$ and $m_3$ is denoted with $a_2$, and the semimajor axis of the `superorbit' of the two bound pairs is denoted with $a_3$.

\subsection{Secular dynamics}
\label{sect:methods:secular}
In \citet{2015MNRAS.449.4221H}, the equations describing the secular evolution of hierarchical four-body systems were given for both the `2+2' and `3+1' configurations. Here, we used the updated code \textsc{SecularMultiple} \citep{2016MNRAS.459.2827H} interfaced in the \textsc{AMUSE} framework \citep{2013CoPhC.183..456P,2013A&A...557A..84P}.  The \textsc{SecularMultiple} code applies to arbitrary hierarchies and numbers of bodies; evidently, its underlying algorithm reduces to the same equations for four bodies. The method is based on an expansion of the Hamiltonian in terms of ratios of the orbital separations (ratios of all combinations of small to large separations), which are assumed to be small. Subsequently, the Hamiltonian is averaged assuming Keplerian motion, and the resulting equations of motion for the orbital vectors $\ve{e}_k$ and $\ve{h}_k$ for all orbits $k$ are integrated numerically. Note that unlike in \citet{2015MNRAS.449.4221H}, \textsc{SecularMultiple} includes the `cross' terms at the octupole order that depend on all three orbits simultaneously, and which were not included in the integrations of \citet{2015MNRAS.449.4221H}. Here, in addition to including all terms at the octupole order, we also included terms corresponding to pairwise binary interactions up and including the fifth order in the separation ratios. Relativistic corrections were included to the first post-Newtonian (PN) order, neglecting terms in the PN potential associated with interactions between binaries.

\subsection{Tidal evolution}
\label{sect:methods:tides}
We included tidal evolution for the planet and stars directly orbiting other bodies. For the `2+2' configuration, tides were taken into account in bodies 0 and 1 in orbit 1, and bodies 2 and 3 in orbit 2 (cf. \F\,\ref{fig:configurations.eps}). For the `3+1' configuration, tides were taken into account only for bodies 0 and 1 in orbit 1. For tidal evolution, we only considered two-body interactions, i.e. we neglected perturbations of other bodies on the direct tidal evolution. Evidently, indirect coupling is taken into account by the changing of the eccentricities due to secular evolution.

We adopted the equilibrium tide model of \citet{1998ApJ...499..853E}, also taking into account the spin evolution (magnitude and direction) of all bodies involved in tidal evolution. For the planet-hosting star, we assumed zero initial obliquities. We included the effect of precession of the orbits due to tidal bulges and rotation. 

The equilibrium tide model is described in terms of the viscous time-scale $t_\mathrm{V}$ (or equivalent related time-scales), the apsidal motion constant $k_\mathrm{AM}$, the gyration radius $r_\mathrm{g}$ and the initial spin period $P_\mathrm{s}$. Our assumed values for the planet and stars, if applicable, are given in Table \ref{table:IC}. Most of the values are adopted from \citet{2007ApJ...669.1298F}. For all tidal interactions, we assumed a constant tidal viscous time-scale $t_\mathrm{V}$ during the simulations. For high-$e$ migration, the viscous time-scale of the planet, $t_\mathrm{V,1}$, is the most important quantity. Apart from its simplicity, a temporally constant $t_\mathrm{V}$ during high-$e$ migration follows from the equations of motion with a number of physically-motivated assumptions \citep{2012arXiv1209.5723S}.

\begin{figure}
\center
\includegraphics[scale = 0.45, trim = 10mm 0mm 0mm 0mm]{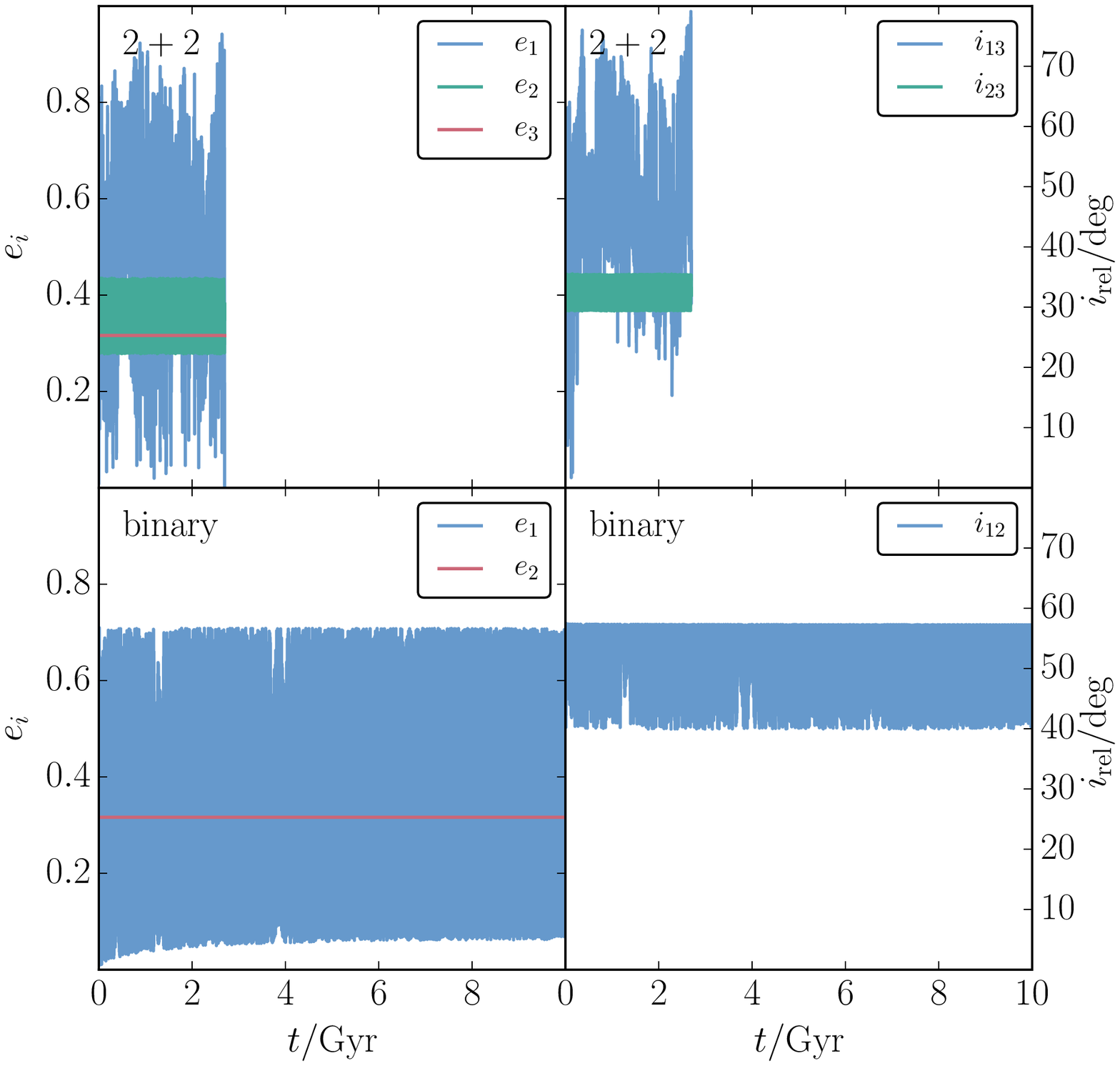}
\includegraphics[scale = 0.45, trim = 10mm 0mm 0mm 0mm]{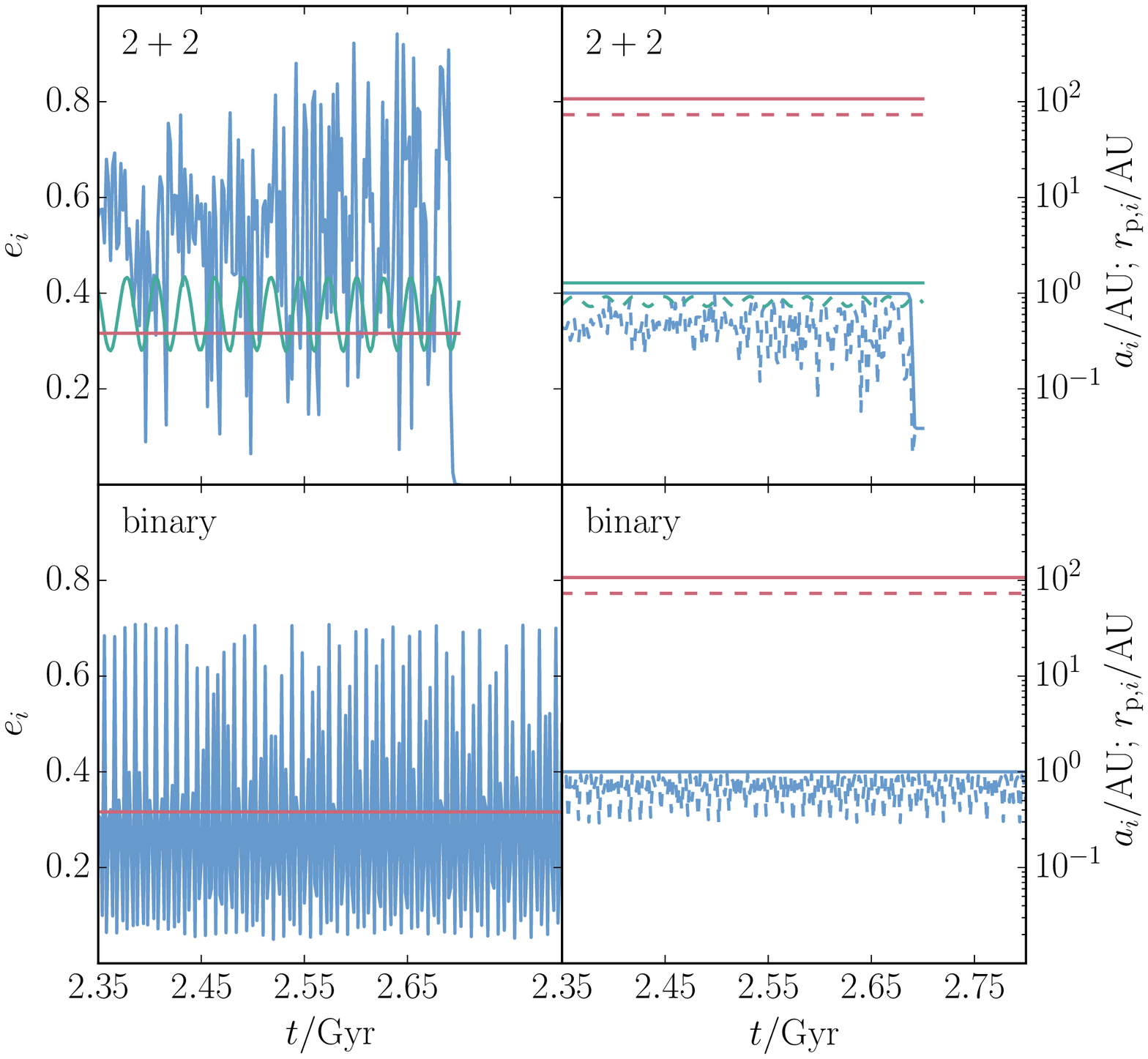}
\caption{\small Example evolution of a system in which a HJ is formed through high-$e$ migration in a stellar triple system (the `2+2' configuration, cf. \F\,\ref{fig:configurations.eps}), whereas in the equivalent binary system, no HJ is formed (cf. \S\,\ref{sect:example}). In the top four panels, the eccentricities $e_i$ (left column) and inclinations relative to the outer orbit (right column) are shown as a function of time (refer to the legends). The first (second) row corresponds to the triple (binary) configuration. In the bottom four panels, a zoom-in is shown around the time at which the HJ is formed in the triple configuration. Eccentricities are shown in the left column; in the right column, solid lines show the semimajor axes and dashed lines show the pericenter distances. The colors in the bottom four panels have the same meaning as in the left column in the top four panels. }
\label{fig:example1_MC03longd.eps}
\end{figure}

\section{Example of differences in secular evolution in stellar triples versus stellar binaries}
\label{sect:example}
Before proceeding with the population synthesis study in \S\,\ref{sect:pop_syn}, we here briefly demonstrate some of the differences in the secular evolution of planets in stellar triple systems compared to stellar binaries. In particular, we consider a planet in a triple in the `2+2' configuration in which the orbit of the planet is excited to high eccentricity, and subsequently a HJ is formed. In contrast, in the equivalent case of a stellar binary, i.e. with orbit 2 replaced by a point mass, the eccentricity of the orbit of the planet is less excited, and no HJ is formed (within 10 Gyr). 

For the example system, we selected a system forming a HJ in the `2+2' triple configuration in the population synthesis of \S\,\ref{sect:pop_syn}. 
The parameters are $m_0 = 1\,\mathrm{M}_\odot$, $m_1 = 1\,M_\mathrm{J}$, $m_2 \approx 0.80\,\mathrm{M}_\odot$, $m_3 \approx 0.42 \, \mathrm{M}_\odot$; $a_1 = 1\,\mathrm{AU}$, $a_2 \approx 1.28\,\mathrm{AU}$, $a_3 \approx 107\,\mathrm{AU}$; $e_1 = 0.01$, $e_2 = 0.31$, $e_3 = 0.32$; $i_1 = 0^\circ$, $i_2 \approx 23.4^\circ$, $i_3 \approx 57.4^\circ$; $\omega_1 \approx 174^\circ$, $\omega_2 \approx 344^\circ$, $\omega_3\approx 246^\circ$; $\Omega_1 \approx 9.70^\circ$, $\Omega_2 \approx 83.1^\circ$, $\Omega_3\approx 92.8^\circ$ and $t_\mathrm{V,1} \approx 0.014\,\mathrm{yr}$. The other parameters are given in Table\,\ref{table:IC}. Note that for these initial parameters, the initial inclination between orbits 1 and 3 is $i_{13} \approx 57.4^\circ$, and $i_{23} \approx 34.5^\circ$ for orbits 2 and 3. 

In the equivalent binary configuration, all orbital parameters associated with orbit 2 are removed, and they are replaced by the parameters of orbit 3 in the triple configuration. The binary companion mass, $\tilde{m}_2$ (here the tilde indicates the binary case), is now given by the sum of the masses within orbit 2 in the triple configuration, i.e. $\tilde{m}_2 = m_2 + m_3 \approx 1.22 \, \mathrm{M}_\odot$. Effectively, orbit 2 in the triple configuration is then reduced to a point mass.

In \F\,\ref{fig:example1_MC03longd.eps}, we show the time evolution of the eccentricities, relative inclinations and semimajor axes for the triple and binary cases. In the binary case, the initial relative inclination of the orbit of the planet with respect to its `outer' orbit 2, is $\tilde{i}_{12} \approx 57.4^\circ$. The maximum eccentricity induced in the former orbit, $\tilde{e}_{1,\mathrm{max}} \sim 0.7$, is consistent with the simplest, lowest-order expectation, $[1-(5/3) \cos(\tilde{i}_{12})^2]^{1/2} \approx 0.72$. This maximum eccentricity is not high enough to induce strong tidal dissipation in orbit 1 (the semimajor axis $\tilde{a}_1$ remains constant), nor trigger a tidal disruption. As expected, the relative inclination $\tilde{i}_{13}$ oscillates between the initial value and $\approx 40^{\circ}$. 

If the stellar companion to $m_0$ is replaced by a binary, then the evolution of orbit 1 is very different. Instead of regular LK oscillations, $e_1$ oscillates much more irregularly. Whereas the maximum inclination between orbit 1 and its outer orbit was limited to the initial value in the binary case, here this maximum inclination is much higher, reaching values of up to $\approx 80^\circ$. The maximum eccentricities reached in orbit 1 are much higher, exceeding 0.9. At an age of $\approx 2.7 \,\mathrm{Gyr}$, $e_1$ becomes high enough for tidal dissipation to become important, and a HJ is rapidly formed. 

These differences between the binary and triple cases can be ascribed to the nodal precession of orbit 3 induced by LK oscillations acting between orbits 2 and 3. This nodal precession can enhance the LK oscillations between orbits 1 and 3 if the ratio of the LK time-scales of associated with the orbital pairs (1,3) and (2,3) are commensurate. The latter condition can be quantified by considering the ratio of LK time-scales as in \citet{2016MNRAS.459.2827H}, i.e. 
\begin{align}
\label{eq:R_2p2}
\mathcal{R}_{2+2} \equiv \frac{ P_\mathrm{LK,13}}{P_\mathrm{LK,23}} \approx \left ( \frac{a_2}{a_1} \right )^{3/2} \left( \frac{m_0+m_1}{m_2+m_3} \right)^{3/2},
\end{align}
where we neglected factors of order unity in estimating the LK time-scales. If $\mathcal{R}_{2+2}\ll 1$, orbit 2 can be considered effectively a point mass, i.e. the binary nature of this orbit can be neglected (see also Section 2.4 of \citealt{2016MNRAS.459.2827H}). In the intermediate regime of $\mathcal{R}_{2+2} \sim 1$, the evolution is generally complex and chaotic as demonstrated in \F\,\ref{fig:example1_MC03longd.eps}, and potentially very high eccentricities can be attained in orbit 1. For the system in \F\,\ref{fig:example1_MC03longd.eps}, $\mathcal{R}_{2+2}$ is indeed close to unity, i.e. $\mathcal{R}_{2+2} \approx 1.1$. 

The rich dynamics of the regime $\mathcal{R}_{2+2} \sim 1$ will be explored in more detail in a later paper.

\section{Population synthesis}
\label{sect:pop_syn}
Using the methods described in \S\,\ref{sect:methods}, we carried out a population synthesis of planets in hierarchical stellar triple systems, considering two distinct hierarchical configurations (cf. panels 1 and 2 in \F\,\ref{fig:configurations.eps}). For reference, and in order to investigate the differences, we also carried out simulations of a planet in a stellar binary mimicking previous studies of high-$e$ migration in stellar binaries \citep{2003ApJ...589..605W,2007ApJ...669.1298F,2012ApJ...754L..36N,2015ApJ...799...27P,2016MNRAS.456.3671A,2016ApJ...829..132P}. With some approximations, high-$e$ migration fractions in stellar binaries can also be calculated analytically \citep{2016MNRAS.460.1086M}.

Consistent with formation beyond the ice line, the planet was initially placed on a nearly circular orbit ($e_1=0.01$) with a semimajor axis $a_1$ ranging between 1 and 4 AU around either the tertiary star (in the case of the `2+2' configuration) or the primary star (in the case of the `3+1' configuration). In the case of a stellar binary, the planet was placed on an orbit around the primary star (S-type orbit). Subsequently, we followed the secular dynamical evolution of the system, also taking into account tidal evolution.

\subsection{Initial conditions}
\label{sect:pop_syn:IC}
For each of the three configurations, we selected three different values of the viscous time-scale $t_\mathrm{V,1}$ of the planet and for each combination, we generated $N_\mathrm{MC}=4000$ systems through Monte Carlo sampling. The viscous time-scales considered were $t_\mathrm{V,1} \approx 0.014,0.14$ and 1.4 yr, respectively. For gas giant planets and high-$e$ migration, \citet{2012arXiv1209.5724S} provided the constraint $t_\mathrm{V,1} \lesssim 1.2 \, \times 10^{4} \, \mathrm{hr} \approx 1.4 \, \mathrm{yr}$, by requiring that a HJ at 5 d is circularized in less than 10 Gyr. Our largest value of $t_\mathrm{V,1}$ corresponds to this value; the other values correspond to 10 and 100 more efficient tidal dissipation. Note that $t_\mathrm{V,1} \approx 1.4\,\mathrm{yr}$ corresponds to a tidal quality factor of $Q_1\approx 1.1\times 10^5$ (cf. equation 37 from \citealt{2012arXiv1209.5724S}), and $Q_1 \propto t_\mathrm{V,1}$.

We made the following assumptions in the Monte Carlo sampling. The mass and radius of the planet were set to $m_1 = 1 \, M_\mathrm{J}$ and $R_1 = R_\mathrm{J}$, respectively. The mass and radius of the star hosting the planet were set to $m_0 = 1\,\mathrm{M}_\odot$ and $R_0 = 1\,\mathrm{R}_\odot$, respectively. The masses of the companion stars, $m_2$ and $m_3$, were computed from $m_2 = q_2 m_0$ and $m_3 = q_3 m_0$, where $q_2$ and $q_3$ were sampled from flat distributions between 0.1 and 1.0. The radii of the companion stars, $R_2$ and $R_3$ (of importance for stellar tides and collisions), were computed using the approximate mass-radius relation $R_i = (m_i/\mathrm{M}_\odot)^{0.8} \, \mathrm{R}_\odot$ for $i\in \{2,3\}$ \citep{1994sse..book.....K}. 

The initial orbit of the planet around its host star was assumed to have an eccentricity $e_1 = 0.01$ and a semimajor axis $a_1$ between 1 and 4 AU, sampled from a flat distribution. The orbital periods $P_2$ and $P_3$ were sampled from a normal distribution in $\mathrm{log}_{10}(P_i/\mathrm{d})$ with a mean of 5.03 and width of 2.28 \citep{2010ApJS..190....1R}, and lower and upper limits of 1 and $10^{10}$ days, respectively. The semimajor axes $a_2$ and $a_3$ were computed from these orbital periods and the sampled masses using Kepler's law. The eccentricities $e_2$ and $e_3$ were computed from a Rayleigh distribution with an rms width of 0.33 between 0.01 and 0.9, approximating the distributions found by \citet{2010ApJS..190....1R}.

In the case of `2+2' systems, sampled orbital combinations were rejected if $a_3/a_2$ was smaller than the critical value for dynamical stability according to the criterion of \citet{2001MNRAS.321..398M}. Here, we treated orbits 2 and 3 as an isolated triple system with the mass of the `tertiary' given by $m_0+m_1$, which is likely a good approximation given the small relative mass of the planet. Also, we required that $a_3/a_1$ be larger than the critical value for dynamical stability according to the criterion of \citet{1999AJ....117..621H}, computed treating the stars orbiting outside the planet as a single body with mass $m_2+m_3$. 

Similarly, in the case of `3+1' systems, stability of orbits 2 and 3 was assessed using the criterion of \citet{2001MNRAS.321..398M}, treating orbit 1 as a point mass with mass $m_0+m_1$. For the planetary orbit, we again applied the criterion of \citet{1999AJ....117..621H}, this time treating orbits 1 and 2 as an isolated system (i.e. neglecting the mass $m_3$).

For the reference case of a stellar binary, only the \citet{1999AJ....117..621H} criterion was applied to orbits 1 and 2. 

Without loss of generality, the inclination of orbit 1 was set to $i_1 = 0^\circ$; the other inclinations $i_2$ and $i_3$ were sampled from flat distributions in $\cos(i_i)$. The arguments of pericenter $\omega_i$ and the longitudes of the ascending nodes $\Omega_i$ were sampled from flat distributions between $0^\circ$ and $360^\circ$. These choices correspond to completely random orbital orientations.

For the other fixed initial parameters, we refer to Table\,\ref{table:IC}. Regarding the stars, we adopted a constant viscous time-scale of $t_\mathrm{V,\star} = 5 \, \mathrm{yr}$, following \citet{2007ApJ...669.1298F}. Assuming a tidal frequency of 4 d, this corresponds to a tidal quality factor of $Q_\star \sim 6 \times 10^8$ or $Q'_\star \equiv 3Q/(4 k_\mathrm{AM,\star}) \sim 3 \times 10^6$, which is typical for Solar-type stars \citep{2007ApJ...661.1180O}.

\begin{figure}
\center
\includegraphics[scale = 0.48, trim = 10mm 0mm 0mm 0mm]{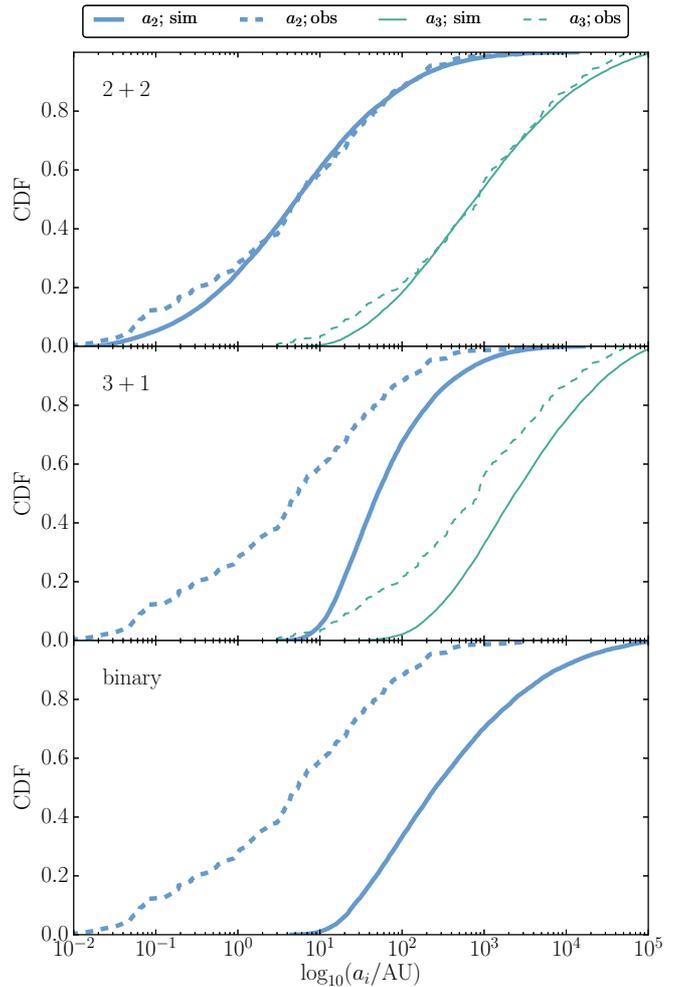}
\caption{\small The initial distributions of the semimajor axes for the three configurations (cf. \F\,\ref{fig:configurations.eps}). Thick blue (thin green) solid lines indicate the sampled distributions of $a_2$ ($a_3$); each row corresponds to a different configuration (note that for a stellar binary, or 3 bodies, only $a_2$ applies). Dashed lines show distributions from observed F- and G-type stellar triples from the sample of \citet{2014AJ....147...87T}, selecting systems in which both the masses and orbital periods are known. }
\label{fig:initial_sma_distributions_combined_MC03long.eps}
\end{figure}

In \F\,\ref{fig:initial_sma_distributions_combined_MC03long.eps}, we show the initial distributions of the semimajor axes for the three configurations. Thick blue (thin green) solid lines indicate the sampled distributions of $a_2$ ($a_3$); each row corresponds to a different configuration (note that for a stellar binary, or 3 bodies, only $a_2$ applies). Dashed lines show distributions from observed F- and G-type stellar triples from the sample of \citet{2014AJ....147...87T}, rejecting systems in which either the masses or orbital periods are unknown. 

For `2+2' systems, the distributions generated for the simulations approximate the observed distributions from \citet{2014AJ....147...87T}, except for small values of $a_2$, i.e. $a_2\sim 10^{-1}\,\mathrm{AU}$. This is because there is an enhancement of short orbital periods in the observations which is believed to be due to LK oscillations induced by the tertiary star combined with tidal dissipation \citep{2007ApJ...669.1298F}. Note that for the `2+2' configuration, tidal dissipation is included the stellar binary (orbit 2), hence the latter effect is taken into account in the simulations. 

For the `3+1' systems, the sampled distributions are distinctly different from the observations, especially regarding $a_2$. This is due to the imposed requirement of stability of the system with the added planet orbiting one of the stars in the inner binary of the stellar triple, pushing $a_2$ to larger values, and, subsequently, $a_3$ as well. Presumably, this is also reflected in the true distributions of the semimajor axes of stellar triples with embedded planets, which are so far very poorly constrained given the small number of observed systems. 

Lastly, in the case of stellar binary systems (cf. the bottom row of \F\,\ref{fig:initial_sma_distributions_combined_MC03long.eps}), there is a similar cutoff in $a_2$ imposed by dynamical stability.

\subsection{Stopping conditions}
\label{sect:pop_syn:SC}
The integrations were stopped if one of the following conditions was met.
\begin{enumerate}
\item The system reached an age of 10 Gyr.
\item A HJ was formed and no further evolution is expected within a Hubble time, i.e. $P_1 < 10 \, \mathrm{d}$ and $e_1 < 10^{-3}$. 
\item The innermost planet was tidally disrupted by its host star, i.e. $r_\mathrm{p,1}=a_1(1-e_1)<r_\mathrm{t}$, where $r_\mathrm{t}$ is given by
\begin{align}
\label{eq:r_t}
r_\mathrm{t} = \eta R_1 \left (\frac{ m_0 }{m_1} \right )^{1/3}.
\end{align}
Here, $\eta$ is a dimensionless parameter; throughout, we assumed $\eta=2.7$ \citep{2011ApJ...732...74G}.
\item The orbit of the planet around its host star became dynamically unstable because of perturbations by the other stars, evaluated using the criterion of \citet{1999AJ....117..621H}.
\item The stellar orbits became dynamically unstable according to the criterion of \citet{2001MNRAS.321..398M}. Note that this only applies to stellar triples.
\item In the `2+2' case, stars collided in orbit 2. In principle, the collision remnant could subsequently secularly excite the orbit of the planet producing a HJ. However, the probability of stellar collision is small (see below).
\item The run time of the simulation exceeded 12 CPU hrs (imposed for practical reasons).
\end{enumerate}

\definecolor{Gray}{gray}{0.9}
\begin{table*}
\scriptsize
\begin{tabular}{ccccccccccccccccccccc}
\toprule
& & \multicolumn{3}{c}{$f_\mathrm{no\, migration}$} & \multicolumn{3}{c}{$f_\mathrm{HJ}$} & \multicolumn{3}{c}{$f_\mathrm{TD}$} & \multicolumn{3}{c}{$f_{\mathrm{DI,planet}}$} & \multicolumn{3}{c}{$f_\mathrm{col,stars}$} & \multicolumn{3}{c}{$f_\mathrm{run\,time\,exceeded}$} \\
\cmidrule(l{2pt}r{2pt}){3-5} \cmidrule(l{2pt}r{2pt}){6-8} \cmidrule(l{2pt}r{2pt}){9-11}  \cmidrule(l{2pt}r{2pt}){12-14} \cmidrule(l{2pt}r{2pt}){15-17} \cmidrule(l{2pt}r{2pt}){18-20}
& & \multicolumn{2}{c}{$t_\mathrm{end} / \mathrm{Gyr}$} & & \multicolumn{2}{c}{$t_\mathrm{end} / \mathrm{Gyr}$} & & \multicolumn{2}{c}{$t_\mathrm{end} /\mathrm{Gyr}$} & & \multicolumn{2}{c}{$t_\mathrm{end} / \mathrm{Gyr}$} & & \multicolumn{2}{c}{$t_\mathrm{end} / \mathrm{Gyr}$} & & \multicolumn{2}{c}{$t_\mathrm{end} / \mathrm{Gyr}$}  \\
\cmidrule(l{2pt}r{2pt}){3-4} \cmidrule(l{2pt}r{2pt}){6-7} \cmidrule(l{2pt}r{2pt}){9-10}  \cmidrule(l{2pt}r{2pt}){12-13} \cmidrule(l{2pt}r{2pt}){15-16} \cmidrule(l{2pt}r{2pt}){18-19}
$1.37 \times 10^{-2}$ & $2+2$ & 0.782 & 0.771 & 0.787 & 0.040 & 0.046 & 0.038 & 0.155 & 0.158 & 0.153 & 0.000 & 0.000 & 0.000 & 0.015 & 0.015 & 0.015 & 0.002 & 0.004 & 0.003  \\
$1.37 \times 10^{-2}$ & $3+1$ & 0.324 & 0.300 & 0.338 & 0.039 & 0.045 & 0.036 & 0.399 & 0.403 & 0.393 & 0.226 & 0.228 & 0.222 & 0.000 & 0.000 & 0.000 & 0.010 & 0.023 & 0.009  \\
$1.37 \times 10^{-2}$ & $\mathrm{binary}$ & 0.817 & 0.805 & 0.820 & 0.033 & 0.037 & 0.031 & 0.145 & 0.146 & 0.143 & 0.001 & 0.001 & 0.001 & 0.000 & 0.000 & 0.000 & 0.005 & 0.013 & 0.006  \\
\midrule
$1.37 \times 10^{-1}$ & $2+2$ & 0.788 & 0.772 & 0.793 & 0.015 & 0.021 & 0.016 & 0.174 & 0.182 & 0.169 & 0.000 & 0.000 & 0.000 & 0.013 & 0.013 & 0.013 & 0.002 & 0.005 & 0.003  \\
$1.37 \times 10^{-1}$ & $3+1$ & 0.326 & 0.303 & 0.340 & 0.019 & 0.023 & 0.016 & 0.423 & 0.429 & 0.415 & 0.217 & 0.218 & 0.213 & 0.000 & 0.000 & 0.000 & 0.014 & 0.026 & 0.014  \\
$1.37 \times 10^{-1}$ & $\mathrm{binary}$ & 0.839 & 0.827 & 0.840 & 0.013 & 0.015 & 0.012 & 0.143 & 0.144 & 0.141 & 0.000 & 0.000 & 0.000 & 0.000 & 0.000 & 0.000 & 0.005 & 0.014 & 0.006  \\
\midrule
$1.37 \times 10^{0}$ & $2+2$ & 0.796 & 0.781 & 0.802 & 0.003 & 0.005 & 0.003 & 0.179 & 0.186 & 0.172 & 0.000 & 0.000 & 0.000 & 0.015 & 0.015 & 0.015 & 0.002 & 0.007 & 0.003  \\
$1.37 \times 10^{0}$ & $3+1$ & 0.332 & 0.305 & 0.346 & 0.002 & 0.004 & 0.002 & 0.444 & 0.450 & 0.433 & 0.208 & 0.210 & 0.202 & 0.000 & 0.000 & 0.000 & 0.012 & 0.028 & 0.014  \\
$1.37 \times 10^{0}$ & $\mathrm{binary}$ & 0.828 & 0.820 & 0.831 & 0.002 & 0.003 & 0.002 & 0.164 & 0.167 & 0.162 & 0.000 & 0.000 & 0.000 & 0.000 & 0.000 & 0.000 & 0.005 & 0.011 & 0.005  \\
\bottomrule
\end{tabular}
\caption{ Outcomes of the $N_\mathrm{MC}=4000$ Monte Carlo realizations for various combinations of the planetary viscous time-scale $t_\mathrm{V,1}$ (in units of yr, rounded to two decimal places), and the three hierarchical configurations (HCs; cf. \F\,\ref{fig:configurations.eps}): a planet orbiting the tertiary star in a stellar triple (`$2+2$'), a planet orbiting one of the stars in the inner binary of a stellar triple (`$3+1$'), and a planet orbiting a star in an S-type orbit in a stellar binary (`binary'). }
\label{table:MC_grid_results}
\end{table*}

\subsection{Results}
\label{sect:pop_syn:results}
\subsubsection{Outcome fractions}
\label{sect:pop_syn:results:fractions}
In Table\,\ref{table:MC_grid_results}, fractions are given of several pathways in the simulations. We distinguish between no migration, i.e. the planet's semimajor axis not changing significantly, the formation of a HJ, the planet being tidally disrupted (`TD'), the planet becoming dynamically unstable (`DI, planet'), stars colliding in orbit 2 (this applies only to the `2+2' cases), and the exceeding of the CPU run time. Each row corresponds to a specific hierarchical configuration (`HC') and viscous time-scale of the planet, $t_\mathrm{V,1}$. The fractions are given after 5 and 10 Gyr of evolution, and by picking a random time between 0.1 and 10 Gyr for each of the $N_\mathrm{MC}=4000$ simulations. 

Table\,\ref{table:MC_grid_results} shows that the HJ fractions between the various configurations are very similar. Nevertheless, there is a systematic trend, for all viscous time-scales, that the HJ fractions for the stellar triple `2+2' and `3+1' configurations are higher compared to the stellar binary configuration. This confirms our expectation of \S\,\ref{sect:introduction}, although the enhancement is only small, not exceeding a few per cent relative to the $N_\mathrm{MC}=4000$ sampled systems. 

A striking difference between the `3+1' and the other configurations is the high fraction of tidally disrupted systems. This can be attributed to smaller values of $a_2$ in simulations with the `3+1' configuration compared to the `outer' orbits of orbit 1 in other configurations (cf. \F\,\ref{fig:initial_sma_distributions_combined_MC03long.eps}), which results in a stronger excitation of the eccentricity of the orbit of the planet, orbit 1, by orbit 2 (see also \S\,\ref{sect:pop_syn:results:triple_orb}). Another difference of the `3+1' configuration compared to the others, is a large fraction of systems in which the planet becomes dynamically unstable, of up to $\sim 0.2$. This can be attributed to enhanced eccentricity of orbit 2 excited by LK oscillations of orbit 3. There are a number of outcomes of such a dynamical instability; this was investigated for a similar situation (in this case with a P-type circumbinary planet, rather than an S-type planet) using $N$-body integrations by \citealt{2016MNRAS.455.3180H}. In the latter work, it was found that ejection of the planet, which is much less massive than the stars, is most likely. In principle, planets could also be tidally captured by stars and forming HJs in this manner (tidal effects were not taken into account in \citealt{2016MNRAS.455.3180H}), although we expect that this is not very likely. Even if a fraction of 0.1 of unstable systems would lead to tidal capture, the enhancement of the HJ fraction would at most be a few per cent. Nevertheless, calculating the contribution of HJs from dynamical instability triggered by secular evolution in these systems is an interesting avenue for future work.

Collisions between stars occur only in the `2+2' configuration, in which the stars labeled $m_2$ and $m_3$ can approach each other at short distances as orbit 2 is driven to high eccentricity by orbit 3. As mentioned above, the collision remnant could subsequently secularly excite the orbit of the planet, producing a HJ. However, the probability of stellar collision is small and therefore also the contribution to forming HJs. 

In a few per cent of systems, the CPU time limit was exceeded and the simulations were stopped (cf. the last column of Table\,\ref{table:MC_grid_results}). These systems are unlikely to produce HJs if the stopping condition had not been imposed; this is discussed in more detail in \S\,\ref{sect:discussion:caveats}.

\subsubsection{HJ orbital period distribution}
\label{sect:pop_syn:results:HJ_orb}

\begin{figure}
\center
\includegraphics[scale = 0.48, trim = 10mm 0mm 0mm 0mm]{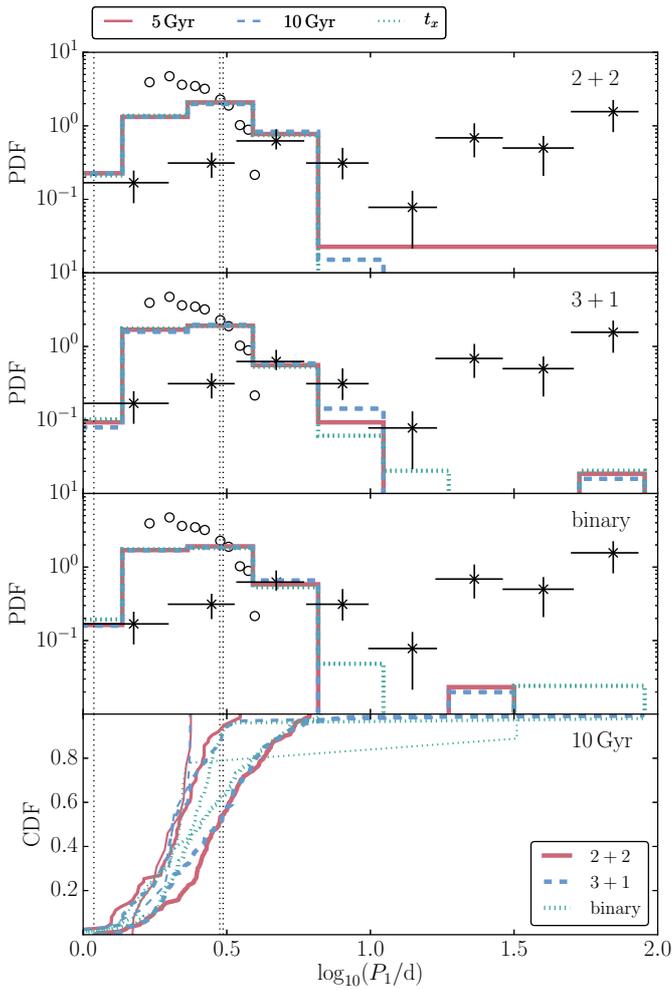}
\caption{\small The orbital period distributions of the HJs formed in the simulations for each configuration (first three panels). In the top three panels, red solid, blue dashed and green dotted lines show the distributions from the simulations after 5 Gyr, 10 Gyr and after a random time (`$t_x$'), combining results from the three viscous time-scales $t_\mathrm{V,1}$. Black crosses with error bars indicate the observations from \citet{2016A&A...587A..64S}, and the orbital periods of the three HJs in stellar triples known to date, WASP-12, HAT-8-P and KELT-4Ab, are indicated with the vertical black dotted lines. Black open circles show the distribution from Fig. 23 of \citet{2016MNRAS.456.3671A} for $m_1 = 1 M_\mathrm{J}$ and $\chi = 100$. Distributions are normalized to unit total area. In the bottom panel, the cumulative distributions are compared for the different configurations at 10 Gyr, and for the three different $t_\mathrm{V,1}$ shown with different line thicknesses (decreasing $t_\mathrm{V,1}$ with increasing line thickness). }
\label{fig:sma_distributions_combined_santerne_MC03long.eps}
\end{figure}

In \F\,\ref{fig:sma_distributions_combined_santerne_MC03long.eps}, we show the orbital period distributions of the short-period planets formed in our simulations for each configuration. In the top three panels, results are combined from the different values of $t_\mathrm{V,1}$. Red solid, blue dashed and green dotted lines show the distributions from the simulations after 5 Gyr, 10 Gyr and after a random time (`$t_x$'). Black crosses with error bars indicate the observations from \citet{2016A&A...587A..64S}, and the orbital periods of the three HJs in stellar triples known to date, WASP-12, HAT-8-P and KELT-4Ab, are indicated with the vertical black dotted lines. For reference, we also include with black open circles the distribution from Fig. 23 of \citet{2016MNRAS.456.3671A} for $M_\mathrm{p} = 1 M_\mathrm{J}$ and $\chi = 100$, where $\chi \equiv 10 \, \tau_1/\mathrm{s}$ and $\tau_1$ is the tidal time lag of the planet (cf. Table 1 of the latter paper). With $k_\mathrm{AM,1} = 0.25$ (cf. Table\,\ref{table:IC}), $m_1 = 1\,M_\mathrm{J}$ and $R_1 = 1 \, R_\mathrm{J}$, $\chi=100$ or $\tau_1 = 10 \, \mathrm{s}$ corresponds to a viscous time-scale of $\approx 0.082 \, \mathrm{yr}$. Distributions are normalized to unit total area. In the bottom panel, the cumulative distributions are compared for the different configurations at 10 Gyr, and for different viscous times-scales $t_\mathrm{V,1}$ shown with different line thicknesses. 

The distributions of the orbital periods are very similar for the different configurations. Compared to the case of a stellar binary and combining the different $t_\mathrm{V,1}$, the KS $D$ and $p$ values are $D\approx 0.10$ and $p\approx 0.29$ for the `2+2' configuration and $D\approx 0.08$ and $p\approx 0.58$ for the `3+1' configuration. These distributions are also similar to those of \citet{2016MNRAS.456.3671A}, although the latter are truncated at a smaller orbital period. 

The dependence on the viscous time-scale $t_\mathrm{V,1}$, shown in the bottom panel of \F\,\ref{fig:sma_distributions_combined_santerne_MC03long.eps}, is easily understood by noting that stronger tides result in larger stalling separations of the HJ (e.g. \citealt{2015ApJ...799...27P}). Another HJ property that depends significantly on $t_\mathrm{V,1}$ is the formation time (cf. \S\,\ref{sect:pop_syn:results:times}). Other properties such as the initial semimajor axes and eccentricities are weakly dependent on $t_\mathrm{V,1}$, and below the corresponding results are shown for all three viscous time-scales combined.

Companions of HJs in the observed sample of \citet{2016A&A...587A..64S} are poorly constrained, and so it is unclear to which extent the HJ orbital period distribution is different for single stars compared to higher-order systems. The orbital periods of the three HJs observed in triple systems so far, indicated with the black vertical dotted lines in \F\,\ref{fig:sma_distributions_combined_santerne_MC03long.eps}, are consistent with the simulations. The period of WASP-12 ($P_1 \approx 1.09\,\mathrm{d}$) lies at the low end of the simulations, whereas the periods of HAT-8-P and KELT-4Ab ($P_1 \approx 3.07$ and $\approx 2.99$ d, respectively) lie at the peak of the simulations around 3 d. 

The simulations fail to produce WJs, i.e. planets with periods between 10 and 100 d, in any meaningful numbers. The number of WJs (out of 4000 systems) at 5 Gyr, 10 Gyr and a random time $t_x$ are 0, 0 and 0 for the `2+2' configuration, 1, 1 and 2 for the `3+1' configuration, and 1, 1, and 2 for the binary configuration, respectively. These few systems appear as low bumps in the orbital period distribution between 10 and 100 days. This is in contrast to observations, which show a much larger number of planets in this period regime compared to HJs. This result is similar to previous studies of high-$e$ migration in stellar binaries (e.g. \citealt{2016ApJ...829..132P,2016arXiv160401781A}), and in multiplanet systems \citep{2016MNRAS.tmp.1471H}.

\subsubsection{Orbital properties of the stellar triple arranged by outcome}
\label{sect:pop_syn:results:triple_orb}

\begin{figure}
\center
\includegraphics[scale = 0.48, trim = 10mm 0mm 0mm 0mm]{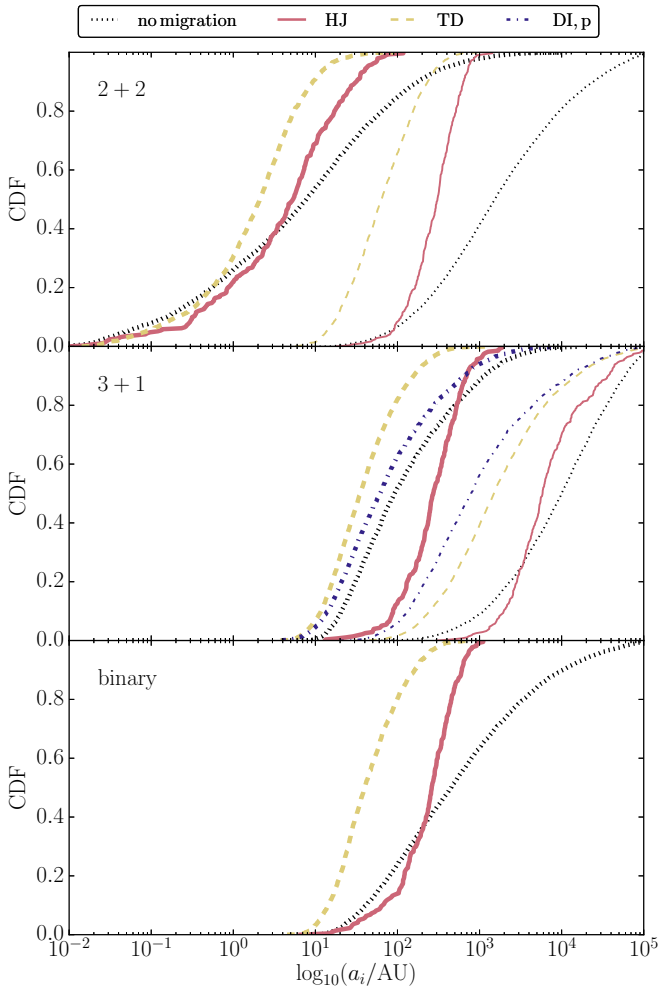}
\caption{\small The distributions of the initial semimajor axes, arranged by various outcomes in the simulations, and combined for the three values of $t_\mathrm{V,1}$. Each panel corresponds to a different configuration; thick (thin) lines apply to orbit 2 (3). Black dotted lines: no migration; red solid lines: HJ formation; yellow dashed lines: planetary tidal disruption; blue dot-dashed lines: dynamical instability of the planet. The initial semimajor axis distributions of all sampled systems were shown with the solid lines in \F\,\ref{fig:initial_sma_distributions_combined_MC03long.eps}. }
\label{fig:arranged_sma_distributions_combined_MC03long.eps}
\end{figure}

In Fig\,\ref{fig:arranged_sma_distributions_combined_MC03long.eps}, we show the distributions of the initial semimajor axes, arranged by various outcomes in the simulations. As mentioned above, these distributions for the HJs are weakly dependent on $t_\mathrm{V,1}$ and the results in Fig\,\ref{fig:arranged_sma_distributions_combined_MC03long.eps} are combined for the three viscous time-scales. Thick (thin) lines apply to orbit 2 (3). We distinguish between no migration (black dotted), HJ formation (red solid), planetary tidal disruption (yellow dashed) and dynamical instability of the planet (blue dot-dashed).

Several interesting constraints on the semimajor axes are revealed in \F\,\ref{fig:arranged_sma_distributions_combined_MC03long.eps}. For `2+2' systems, HJs are only formed if $a_2\lesssim 10^2\, \mathrm{AU}$, and $a_3$ needs to be $\gtrsim 20\,\mathrm{AU}$ and $\lesssim 10^3\,\mathrm{AU}$. These limits can be understood as follows. The lower limit on $a_3$ coincides with the lowest sampled value of $a_3$ (cf. the thin red solid and black dotted lines in the top panel of \F\,\ref{fig:arranged_sma_distributions_combined_MC03long.eps}), and which do not lead to tidal disruption. The lowest sampled value of $a_3$ arises from the requirement of dynamical stability of the system (cf. \S\,\ref{sect:pop_syn:IC} and \F\,\ref{fig:initial_sma_distributions_combined_MC03long.eps}). The upper limit on $a_3$ for HJ systems arises from the well-known phenomenon that LK cycles are suppressed if the `outer' orbit (here, orbit 3) is wide compared to the `inner' orbit (here, orbit 1), in which case precession due to GR, tidal bulges and/or rotation quenches LK oscillations (e.g. \citealt{2013ApJ...773..187N}). 

This upper limit on $a_3$, $a_3\lesssim 10^3\,\mathrm{AU}$, translates to an upper limit on $a_2$ because of the requirement of dynamical stability. A typical ratio of $a_3/a_2$ required for dynamical stability in our simulations is $\sim 6$ (assuming the criterion of \citealt{2001MNRAS.321..398M}). This implies an upper limit on $a_2$ of $\sim 10^3 /6 \,\mathrm{AU} \approx 167 \,\mathrm{AU}$, and which is indeed roughly consistent with the upper limit $a_2\lesssim 10^2\, \mathrm{AU}$ found in the simulations. 

Similar limits apply to the tidal disruption systems in the `2+2' configuration. Most of these systems have an initial $a_3$ smaller compared to the HJ systems: for very small $a_3$, close to dynamical instability with respect the planet (orbit 1), strong secular dynamical interactions excite very high eccentricities in orbit 1, leading to tidal disruption of the planet. The upper limit $a_3\lesssim 5\times10^2\,\mathrm{AU}$ is lower compared to HJs systems, implying a lower value of the upper limit on $a_2$, in this case $a_2 \lesssim 50\, \mathrm{AU}$. Similarly to the HJ systems, the upper limits on $a_3$ and $a_2$ are roughly consistent with requiring dynamical stability of orbits 2 and 3, respectively.

Considering `3+1' systems, $a_2$ for HJ systems is larger compared to the `2+2' configuration, which can be ascribed to the different hierarchy, i.e. $a_2\gg a_1$ is always required for dynamical stability for `3+1' systems, whereas this is not the case for `2+2' systems (for the latter, systems with similar $a_1$ and $a_2$ are typically dynamically stable unless $a_3$ is small). The typical values of $a_2$ for HJs in `3+1' systems are very similar to those of $a_3$ for HJs in `2+2' systems. These similar limits can again be explained by the requirement of dynamical stability and the suppression of LK oscillations for large `outer' orbit semimajor axes (i.e. in the `3+1' case, the outer orbit is orbit 2, and in the `2+2' case, the outer orbit is orbit 3). 

For `3+1' planetary tidal disruption systems, $a_2$ is smallest, with $a_2\lesssim 5\times 10^2\,\mathrm{AU}$. For systems in which the planet becomes dynamically unstable, $a_2$ lies roughly in between the typical values for tidal disruption and HJ formation. These instabilities are driven by secular eccentricity excitation of orbit 2 by the torque of orbit 3, i.e. instability would not have occurred in the absence of the fourth body.

The implications of these constraints on $a_2$ and $a_3$ in triple systems are discussed in \S\,\ref{sect:discussion:constraints}. 

\begin{figure}
\center
\includegraphics[scale = 0.48, trim = 10mm 0mm 0mm 0mm]{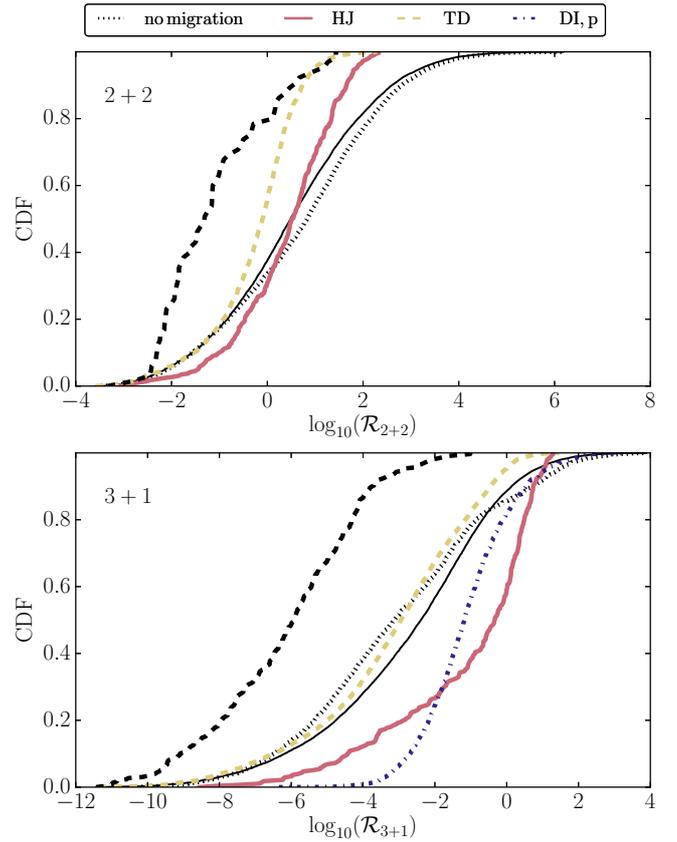}
\caption{\small The distributions of $\mathcal{R}_{2+2}$ (cf. equation~\ref{eq:R_2p2}) and $\mathcal{R}_{3+1}$ (cf. equation~\ref{eq:R_3p1}) shown for `2+2' and `3+1' systems in the top and bottom panels, respectively, and arranged by the outcomes in the simulations (combining results for the three values of $t_\mathrm{V,1}$). The black solid lines show the initial distributions of  $\mathcal{R}_{2+2}$ and  $\mathcal{R}_{3+1}$, and the thick black dashed lines show the initial distributions for the systems in which the run time was exceeded (cf. \S\,\ref{sect:discussion:caveats}). }
\label{fig:R_distributions_combined_MC03long.eps}
\end{figure}

In addition to the semimajor axes, a quantity of interest is the ratio of LK time-scales applied to different orbits, with its definition depending on the hierarchy. For `2+2' systems, the relevant ratio, $\mathcal{R}_{2+2} \equiv P_\mathrm{LK,13}/P_\mathrm{LK,23}$, was introduced in \S\,\ref{sect:example} (cf. equation~\ref{eq:R_2p2}). As mentioned there, particularly high eccentricities, and therefore HJs and tidal disruptions, are expected in the regime $\mathcal{R}_{2+2}\sim 1$. In the top panel of \F\,\ref{fig:R_distributions_combined_MC03long.eps}, the distribution of $\mathcal{R}_{2+2}$ is shown for `2+2' systems, arranged by the outcomes in the simulations (combining results for the three values of $t_\mathrm{V,1}$). HJs systems indeed show a preference for $\mathcal{R}_{2+2}\sim 1$ compared to non-migrating systems. For tidal disruption systems, the distribution is even more concentrated towards $\mathcal{R}_{2+2}\sim 1$, consistent with the notion that more violent eccentricity excitation is more likely to result in tidal disruption.

For `3+1' systems, the relevant LK time-scale ratio is associated with the orbital pairs (1,2) and (2,3) \citep{2015MNRAS.449.4221H}, and is given by
\begin{align}
\label{eq:R_3p1}
\mathcal{R}_{3+1} \equiv \frac{P_{\mathrm{LK},12}}{P_\mathrm{LK,23}} \approx \left ( \frac{ a_2^3}{a_1 a_3^2} \right )^{3/2} \left ( \frac{m_0+m_1}{m_0+m_1+m_2} \right )^{1/2} \frac{m_3}{m_2} \left ( \frac{ 1-e_2^2}{1-e_3^2} \right )^{3/2}.
\end{align}
As shown by \citet{2015MNRAS.449.4221H}, complex eccentricity oscillations and potentially high eccentricities are expected if $\mathcal{R}_{3+1} \sim 1$. In the bottom panel of \F\,\ref{fig:R_distributions_combined_MC03long.eps}, we show the distribution of $\mathcal{R}_{3+1}$ for the `3+1' configuration, arranged by the outcomes. The HJ systems indeed display a preference for larger values of $\mathcal{R}_{3+1}$ close to unity (median value of $\log_{10}\mathcal{R}_{3+1} \sim 0$) compared to the non-migrating systems (median value of $\log_{10}\mathcal{R}_{3+1} \sim -3$). 

The tidally disrupted systems, on the other hand, show a distribution of $\mathcal{R}_{3+1}$ which is biased toward small values of $\mathcal{R}_{3+1}$, seemingly contradicting the above expectation. However, as already mentioned in \S\,\ref{sect:pop_syn:results:fractions}, the fraction of tidally disrupted planets in the `3+1' configuration is large owing to the small value of $a_2$ compared to $a_3$ in the `2+2' configuration, and $a_2$ in the binary configuration (cf. \F\,\ref{fig:initial_sma_distributions_combined_MC03long.eps}). As shown in \F\,\ref{fig:arranged_sma_distributions_combined_MC03long.eps}, $a_2$ for the tidal disruption systems is indeed strongly biased to small values, $10\,\mathrm{AU} \lesssim a_2 \lesssim 5\times10^2\,\mathrm{AU}$, in which case a very high eccentricity can be induced in orbit 1 regardless of orbit 3, leading to tidal disruption. Such small values of $a_2$ correspond to small values of $\mathcal{R}_{3+1}$ (cf. equation~\ref{eq:R_3p1}).

\begin{figure}
\center
\includegraphics[scale = 0.48, trim = 10mm 0mm 0mm 0mm]{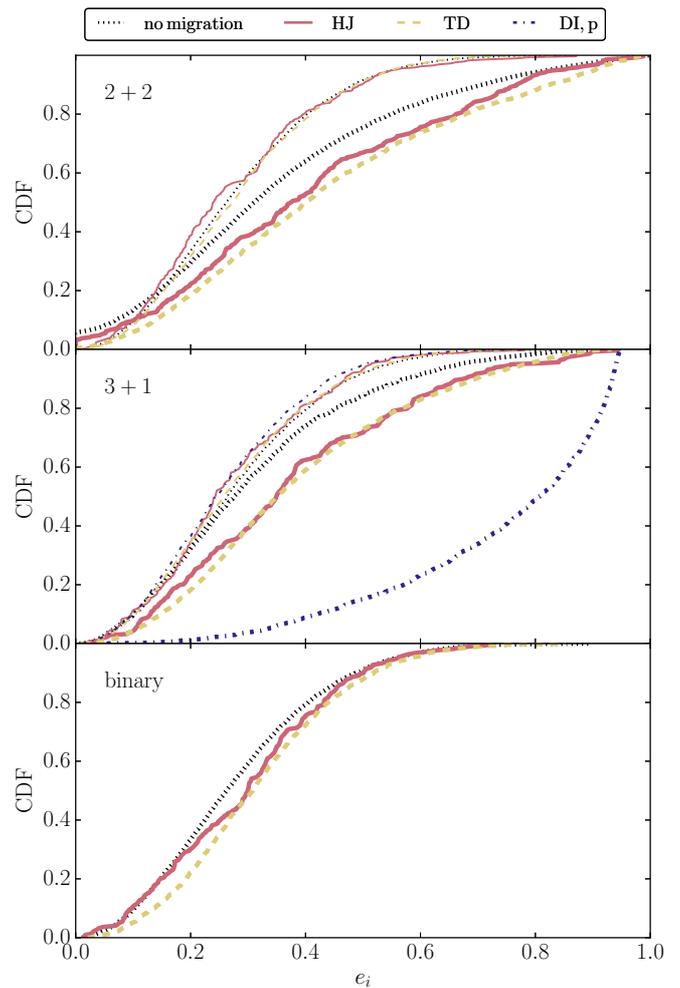}
\caption{\small Similar to \F\,\ref{fig:arranged_sma_distributions_combined_MC03long.eps} now showing the initial distributions of the eccentricities (combining results for the three values of $t_\mathrm{V,1}$). Thick (thin) lines show the distributions for $e_2$ ($e_3$). Note that the initial eccentricities were sampled from Rayleigh distributions, for which the cumulative distribution $\propto 1-\mathrm{exp}(-\beta e_i^2)$. }
\label{fig:arranged_es_distributions_combined_MC03long.eps}
\end{figure}

In Fig\,\ref{fig:arranged_es_distributions_combined_MC03long.eps}, the initial distributions of the eccentricities are shown arranged by the various outcomes, similarly to \F\,\ref{fig:arranged_sma_distributions_combined_MC03long.eps}. As might be expected, there is some preference for HJ and tidal disruption systems to have larger initial $e_3$ (for `2+2' systems) and $e_2$ (for `3+1' systems). The planetary dynamical instability outcome shows a strong preference for high values of $e_2$. If $e_2$ is high initially, only a small change of $e_2$ due to the secular torque of orbit 3 is sufficient to drive instability of the planet.

\subsubsection{HJ formation times}
\label{sect:pop_syn:results:times}

\begin{figure}
\center
\includegraphics[scale = 0.48, trim = 10mm 0mm 0mm 0mm]{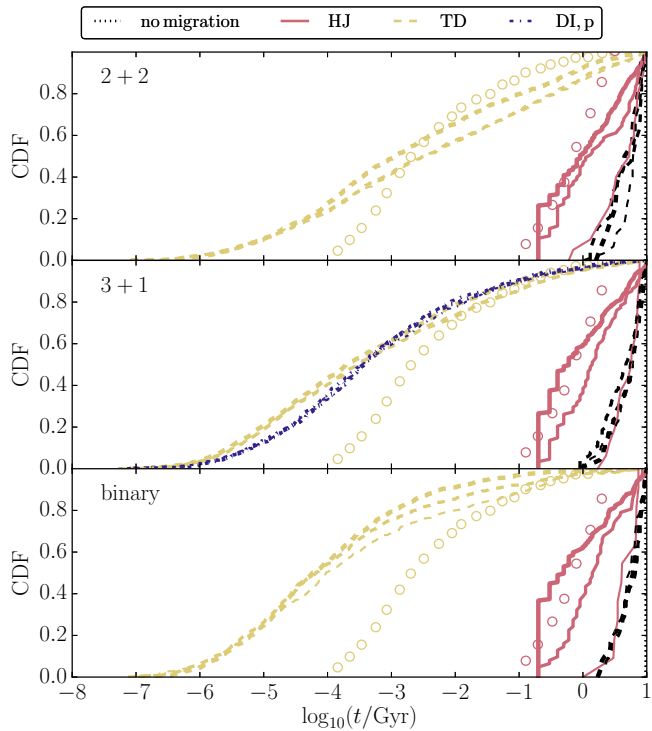}
\caption{\small Distributions of the `end times' associated with various outcomes of the simulations, indicated in the legend. Results are shown separately for the three assumed values of $t_\mathrm{V,1}$ with different line thicknesses (decreasing $t_\mathrm{V,1}$ with increasing line thickness). In addition, the thick black dashed lines show the initial distributions for the systems in which the run time was exceeded (cf. \S\,\ref{sect:discussion:caveats}). Red (yellow) open circles show data for HJs (tidal disruptions) from the second panel of Fig. 22 ($m_1 = 1 M_\mathrm{J}$) of \citet{2016MNRAS.456.3671A}. }
\label{fig:end_times_combined_MC03long.eps}
\end{figure}

The distributions of the `end times' associated with various outcomes of the simulations are shown in \F\,\ref{fig:end_times_combined_MC03long.eps}, each row corresponding to a different configuration. Here, `end time' corresponds to the time of HJ formation or time of disruption for the tidally disrupted planets. Results are shown separately for the three assumed values of $t_\mathrm{V,1}$ with different line thicknesses (decreasing $t_\mathrm{V,1}$ with increasing line thickness).

As expected, the shorter values of $t_\mathrm{V,1}$ (stronger tides) lead to shorter HJ formation times. Similarly to the distribution of the HJ orbital periods, the HJ formation times are not significantly different between the three configurations. The red open circles show data for simulations in stellar binaries from the second panel of Fig. 22 ($m_1 = 1 M_\mathrm{J}$) by \citet{2016MNRAS.456.3671A}, and which are roughly consistent with our simulations. Tidal disruptions, however, occur earlier in our simulations of stellar binaries compared to \citet{2016MNRAS.456.3671A} (cf. the yellow open circles in \F\,\ref{fig:end_times_combined_MC03long.eps}), and this may be because of different assumptions on the orbital parameters. There are no large differences in the tidal disruption time distributions between the different configurations.

\subsubsection{Stellar obliquities}
\label{sect:pop_syn:results:obl}

\begin{figure}
\center
\includegraphics[scale = 0.48, trim = 10mm 0mm 0mm 0mm]{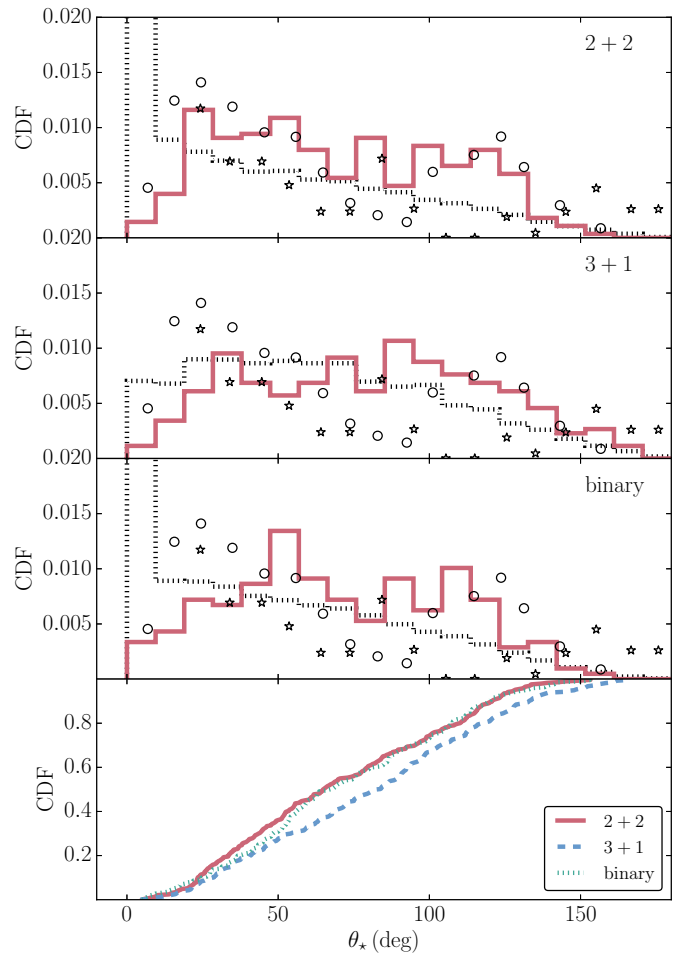}
\caption{\small Top three panels: the distributions of the stellar obliquity for the non-migrating (black dotted lines) and HJ systems (red solid lines) for the three configurations (combining results for the three values of $t_\mathrm{V,1}$). Data from Fig. 24 of \citet{2016MNRAS.456.3671A}  for $m_1 = 1 M_\mathrm{J}$ and $\chi = 100$ are shown with black open circles. Black stars show observational data adopted from \citet{2014PNAS..11112610L}. Cumulative distributions between the three configurations are shown in the bottom panel. }
\label{fig:stellar_obliquity_distributions_MC03long.eps}
\end{figure}

In \F\,\ref{fig:stellar_obliquity_distributions_MC03long.eps}, we show the distributions of the stellar obliquity, i.e. the angle between the spin of the planet-hosting star and the orbit of the planet, for the non-migrating (black dotted lines) and HJ systems (red solid lines). Results are combined for the three values of $t_\mathrm{V,1}$. Data from Fig. 24 of \citet{2016MNRAS.456.3671A}  for $m_1 = 1 M_\mathrm{J}$ and $\chi = 100$ are shown with black open circles. Additionally, we show with black stars observational data adopted from \citet{2014PNAS..11112610L}. The initial obliquities in the simulations were assumed to be zero. 

In the case of a stellar binary, there are two distinct, but broad, peaks in the obliquity distribution near $\sim 50^\circ$ and $\sim 100^\circ$. Such peaks in the obliquity distribution are well known to arise in stellar binaries (\citealt{2007ApJ...669.1298F,2014ApJ...793..137N,2016MNRAS.456.3671A}; see \citealt{2016arXiv160703937S} for a detailed study on the origin of the bimodel distribution). There appears to be a discrepancy with \citet{2016MNRAS.456.3671A}, who find peaks near values of $\sim 25^\circ$ and $\sim 130^\circ$. This may be due various reasons, including the initially shorter spin period of $P_\mathrm{s,\star} = 2.3\,\mathrm{d}$ rather than 10 d for the host star, magnetic braking, which was included in \citet{2016MNRAS.456.3671A} but not in this work (magnetic braking results in the spin down of the star, which, due to spin-orbit coupling, can significantly change the final obliquity, \citealt{2016arXiv160703937S}), or different $t_\mathrm{V,1}$.  

For stellar triples, the obliquity distributions seem marginally less peaked compared to the case of a stellar binary. In the bottom panel of \F\,\ref{fig:stellar_obliquity_distributions_MC03long.eps}, the cumulative distributions between the three configurations are compared directly. Compared to the case of a stellar binary, the KS $D$ and $p$ values are $D\approx 0.06$ and $p\approx 0.66$ for the `2+2' configuration and $D\approx 0.13$ and $p\approx 0.03$ for the `3+1' configuration. The differences in the obliquity distributions in stellar triples compared to binaries might be ascribed to the more chaotic nature of the eccentricity and inclination oscillations in the regime $\mathcal{R}\sim 1$ compared to binaries, in which the regime for chaotic evolution is much smaller (i.e. only for marginally hierarchical systems, e.g. \citealt{2014ApJ...791...86L}). This implies larger possible variations in inclinations at times of maximum eccentricity and therefore less peaked distributions of the obliquity after tidal dissipation has shrunk the planetary orbit.

\subsubsection{Initial inclinations}
\label{sect:pop_syn:results:obl}
\begin{figure}
\center
\includegraphics[scale = 0.46, trim = 5mm 0mm 0mm 0mm]{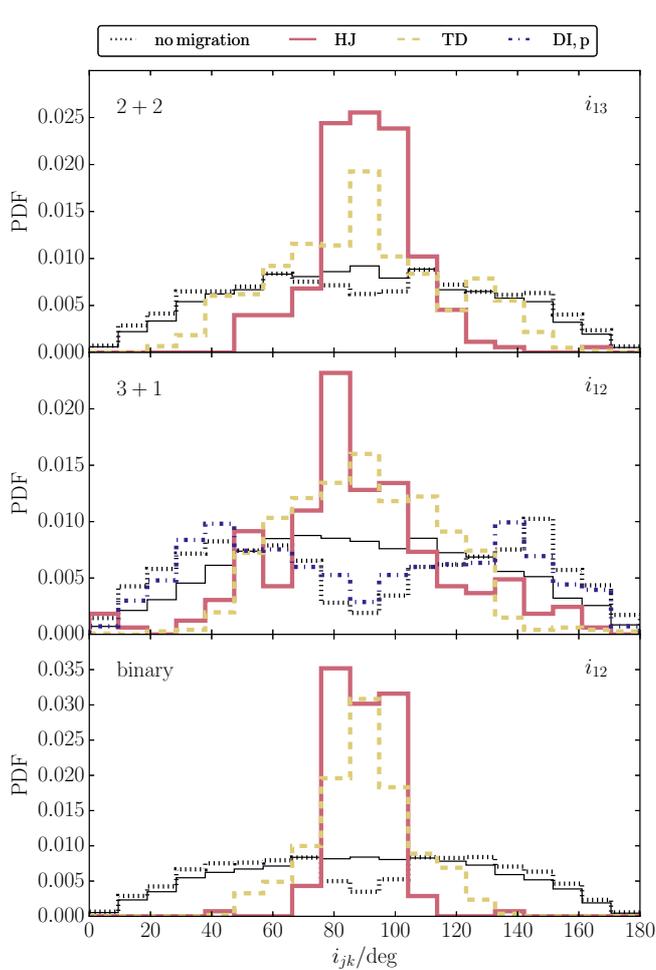}
\caption{\small The distributions of the initial mutual inclination between the planetary orbit and its parent orbit, arranged by the simulation outcomes (combining results for the three $t_\mathrm{V,1}$). This mutual inclination is $i_{12}$ for `3+1' and `binary' systems; for the `2+2' configuration, it is $i_{13}$ (cf. \F\,\ref{fig:configurations.eps}). }
\label{fig:mutual_inclination_distributions_planet_parent_combined_MC05long.eps}
\end{figure}

\begin{figure}
\center
\includegraphics[scale = 0.46, trim = 5mm 0mm 0mm 0mm]{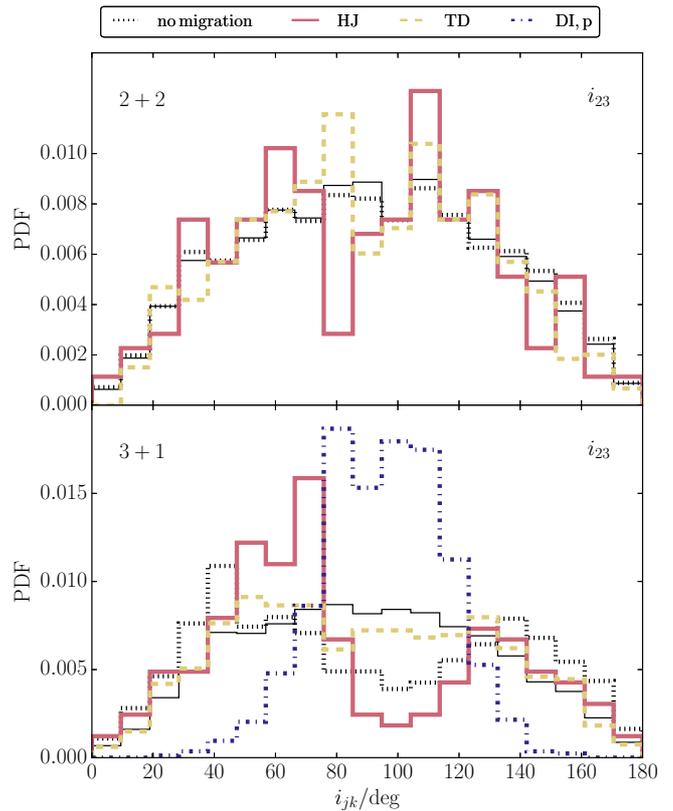}
\caption{\small Similar to \F\,\ref{fig:mutual_inclination_distributions_planet_parent_combined_MC05long.eps}, now showing the distributions of the initial mutual inclination between orbits 2 and 3. }\label{fig:mutual_inclination_distributions_star_parent_combined_MC05long.eps}
\end{figure}

Lastly, we consider the initial inclinations. In \F\,\ref{fig:mutual_inclination_distributions_planet_parent_combined_MC05long.eps}, the distributions of the initial mutual inclination between the planetary orbit and its parent orbit are shown, arranged by the simulation outcomes (combining results for the three $t_\mathrm{V,1}$). This mutual inclination is $i_{12}$ for `3+1' and `binary' systems; for the `2+2' configuration, it is $i_{13}$ (cf. \F\,\ref{fig:configurations.eps}). We recall that the initial orbital orientations were assumed to be random (cf. \S\,\ref{sect:pop_syn:IC}). As expected, high mutual inclinations, near $90^\circ$, are required to excite high eccentricities leading to tidal disruptions or HJs. There are some differences between the triple and binary star cases: for the former, the range of inclinations for tidal disruption and HJ systems around $90^\circ$ is larger, even extending to inclinations lower than $\approx 40^\circ$ or higher than $\approx 130^\circ$. The enlargement of the inclination window for high eccentricities is most pronounced for the `3+1' configuration, for which HJs can be formed even for nearly coplanar systems (prograde or retrograde). For `3+1' configurations, this can be understood by noting that orbits 1 and 2 can be made highly inclined by changes of the absolute inclination of orbit 2 driven by the torque of orbit 3 \citep{2015MNRAS.449.4221H}.

In \F\,\ref{fig:mutual_inclination_distributions_star_parent_combined_MC05long.eps}, we show a similar figure with the distributions of the initial mutual inclination between orbits 2 and 3 (stellar orbits only). This inclination only applies to the triple star cases. For the `2+2' configuration, the tidal disruption and HJ systems show no strong dependence on $i_{23}$, although there is a preference for tidal disruption systems and a lack of HJ systems near $i_{23} = 80^\circ$; the origin for this is unclear. Regarding the `3+1' configuration, systems in which the planet becomes dynamically unstable show a strong preference for high inclinations of $i_{23}$ near $90^\circ$. For such high inclinations, the maximum eccentricity excited in orbit 2 is high, making a dynamical instability with respect to orbit 1 likely. This effect prevents a large fraction of highly inclined systems to become tidal disruption or HJ systems. The latter, instead, show peaks in the initial distribution of $i_{23}$ at lower inclinations near $60^\circ$ and higher inclinations near $130^\circ$.

\section{Discussion}
\label{sect:discussion}

\subsection{Caveats of the simulations}
\label{sect:discussion:caveats}
The systems in our simulations were sampled assuming a planet initially orbiting its host star between $a_1 = 1$ and $4\,\mathrm{AU}$ and assuming lognormal distributions for the stellar orbits, rejecting systems that are dynamically unstable according to (approximate) analytic stability criteria (cf. \S\,\ref{sect:pop_syn:IC}). Although this method excludes systems that are almost certainly dynamically unstable, it may not reflect the intrinsic population of stellar triples with planets, for which the orbital distributions are currently essentially unconstrained. In other words, aside from dynamical stability constraints, it is unclear if the planet formation process(es) that occur in stellar triples with one (or more) planet(s) have other effects on the properties of triples. We believe that the method adopted in this work is warranted, given the lack of observational constraints and better alternatives. Nevertheless, it should be taken into account that the our results depend on the assumed orbital distributions, most importantly the distributions of the semimajor axes and eccentricities, and the assumed viscous time-scales. The latter dependence is shown explicitly in Table\,\ref{table:MC_grid_results}, and in Figs.~\ref{fig:sma_distributions_combined_santerne_MC03long.eps} and \ref{fig:end_times_combined_MC03long.eps}.

It was mentioned in \S\,\ref{sect:pop_syn:results:fractions} that the imposed run time of 12 CPU hrs was exceeded in a few per cent of systems, but that these systems are unlikely to produce HJs if the integrations had not been stopped before reaching an age of 10 Gyr. In these systems, the LK time-scales are extremely short compared to the integration time of 10 Gyr, such that not all cycles could be completed within the CPU time limit. These systems have distinct orbital properties compared to the HJ systems, and are therefore unlikely to produce HJs if the integrations had not been stopped before completion. This is illustrated in Figures \,\ref{fig:R_distributions_combined_MC03long.eps} and \ref{fig:end_times_combined_MC03long.eps}, where the distributions of $\mathcal{R}_{2+2}$ and $\mathcal{R}_{3+1}$ and of the end times, respectively, are shown for the run time exceeded systems with the thick black dashed lines. The ratios $\mathcal{R}_{2+2}$ and $\mathcal{R}_{3+1}$ are typically very small for the run time exceeded systems, implying very short LK time-scales for the innermost pairs (i.e. orbits 1-3 for the `2+2', and orbits 1-2 for the `3+1' configurations). Also, the ages at which the simulations were terminated (cf. \F\,\ref{fig:end_times_combined_MC03long.eps}) are typically longer than the HJ formation times.

\subsection{Comparisons of the HJ fractions to other high-$e$ migration scenarios and observations}
\label{sect:discussion:comparison}
The highest HJ fraction in stellar triples found in our simulations was $\approx 0.05$, compared to $\approx 0.04$ in our fiducial binary star simulations (cf. Table\,\ref{table:MC_grid_results}). The enhancement due to the richer (secular) dynamics therefore turn out to be only small. We note, however, that the fraction of tidal disruptions in `3+1' configurations is markedly higher compared to binaries (up to $\approx 0.5$), mainly due to the planet+star subsystem being orbited by a relatively tight outer orbit. This finding of apparently low HJ fractions is similar to the result of \citet{2016MNRAS.tmp.1471H} for secular eccentricity excitation in multiplanet systems (between 3 and 5 planets orbiting a single star). In the latter work, it was found that, although potentially very high eccentricities can be attained in orbit of the innermost planet, these high eccentricities typically lead to a `fast' tidal disruption, rather than a slower process of orbital tidal decay. 

To compare our simulated HJ fractions to observations, we assume a stellar triple fraction of $f_\mathrm{triple}=0.1$ (e.g. \citealt{2014AJ....147...87T}) and a giant planet fraction in triples of $f_\mathrm{GP} = 0.1$. Adopting our highest fraction of $f_\mathrm{HJ,triple,sim}=0.05$, the expected HJ fraction around Solar-type MS stars due to our scenario of secular high-$e$ `triple' migration is $f_\mathrm{HJ,triple} = f_\mathrm{triple} f_\mathrm{HJ,triple,sim} f_\mathrm{GP} = 5\times 10^{-4}$. In other words, given a Solar-type MS star (either single or multiple), the fraction of these stars around which a HJ is expected to form through our scenario is $\sim 5 \times 10^{-4}$. The observed HJ fraction around Solar-type MS stars is $f_\mathrm{HJ,obs}\sim 0.01$ (e.g. \citealt{2012ApJ...753..160W}), i.e. 20 times larger. Therefore, our `triple' scenario cannot resolve the existing discrepancy that the predicted HJ formation rates of high-$e$ migration in stellar binaries are too low compared to observations. This is because the formation efficiency is only marginally higher, whereas the fraction of stellar triple systems is lower compared to binaries.

\subsection{Constraints on stellar triples hosting HJs and comparisons to observations}
\label{sect:discussion:constraints}
As discussed in \S\,\ref{sect:pop_syn:results:triple_orb}, we find that HJs can only be formed through high-$e$ migration in stellar triples with specific orbital configurations (cf. \F\,\ref{fig:arranged_sma_distributions_combined_MC03long.eps}). The three HJs observed in triples so far, WASP-12b \citep{2009ApJ...693.1920H,2013MNRAS.428..182B,2014ApJ...788....2B}, HAT-P-8b \citep{2009ApJ...704.1107L,2013MNRAS.428..182B,2014ApJ...788....2B}, and KELT-4Ab \citep{2016AJ....151...45E}, orbit the tertiary star of the triple, i.e. the `2+2' configuration applies. For this configuration, we found the constraints $a_2 \lesssim 10^2\,\mathrm{AU}$ and $20\,\mathrm{AU} \lesssim a_3 \lesssim 10^3\,\mathrm{AU}$. 

The projected separations of the stellar binaries in WASP-12b and HAT-P-8 were found by \citet{2014ApJ...788....2B} to be $\approx 21\,\mathrm{AU}$ and $\approx 15\,\mathrm{AU}$, respectively. For simplicity, we adopt these separations as semimajor axes, i.e. $a_2\sim 21\mathrm{AU}$ and $a_2\sim 15\,\mathrm{AU}$ for these systems, respectively. For KELT-4Ab, \citet{2016AJ....151...45E} found a projected separation of $10.3\,\mathrm{AU}$ for the stellar binary, which we adopt as $a_2 \sim 10 \, \mathrm{AU}$. In all three cases, the semimajor axis $a_2$ satisfies $a_2 \lesssim 10^2\,\mathrm{AU}$. 

Assuming an angular separation of $1.047''$ between WASP-12b and its stellar binary companion \citep{2013MNRAS.428..182B} and a distance of 427 pc \citep{2011AJ....141..179C}, the projected separation is $\approx 447 \,\mathrm{AU}$. For HAT-P-8, an angular separation of $1.027''$ between HAT-P-8 and its stellar binary companion \citep{2013MNRAS.428..182B} and a distance of 230 pc \citep{2009ApJ...704.1107L} imply a projected separation of $\approx 236 \,\mathrm{AU}$. \citet{2016AJ....151...45E} found a projected separation for KELT-4Ab of $328\,\mathrm{AU}$. Again adopting the projected separations as semimajor axes, the semimajor axes $a_3$ of these systems satisfy $a_3 \lesssim 10^3\,\mathrm{AU}$.

This shows that, at least approximately given that only the projected separations are known, the observed HJs are consistent with the constraints given by the simulations. Moreover, if a HJ were to be found in a stellar triple with distinctly different orbital parameters (e.g. $a_2 \gg 10^2\,\mathrm{AU}$ for the `2+2' configuration), this would indicate HJ formation through another mechanism such as disk migration.

\section{Conclusions}
\label{sect:conclusions}
We have considered the formation of HJs through high-$e$ migration in stellar triple systems. In this scenario, the orbit of a Jupiter-like planet is excited to high eccentricity by secular perturbations from the stars. The subsequent close pericenter passages lead to tidal decay of the orbit, potentially producing a HJ. We have considered two hierarchical configurations of the planet in the stellar triple system: the planet orbiting the tertiary star (`2+2' configuration; cf. panel 1 of \F\,\ref{fig:configurations.eps}), and the planet orbiting one of the stars in the inner binary (`3+1' configuration; cf. panel 2 of \F\,\ref{fig:configurations.eps}). For reference, we also carried out simulations in stellar binaries, in which case the planet is orbiting one of the stars in an S-type orbit. Our main conclusions are as follows.

\medskip \noindent 1. Although the HJ fraction in stellar triple systems is larger compared to stellar binaries, the enhancement is small, at most a few tens of per cent. In our simulations, the HJ fraction (at an age of 10 Gyr) is largest for the `2+2' systems, i.e. $0.046$ assuming a viscous time-scale of $t_\mathrm{V,1} \approx 0.014\,\mathrm{yr}$ for the planet. The corresponding fraction for `3+1' systems is $0.045$, compared to $0.037$ for stellar binaries. For longer viscous time-scales, the fractions are lower but the trend of slightly larger HJ fractions for triples persists. 

\medskip \noindent 2. The orbital period distributions of the HJs in our simulations are very similar for the different configurations, peaking around 3 d. The orbital periods of the three HJs in stellar triples observed so far are consistent with the orbital period distributions from the simulations. No significant number of WJs is produced in the simulations for any of the configurations. 

\medskip \noindent 3. In our simulations, HJs are formed only for specific ranges of the semimajor axes of the orbits of the triple system (cf. \S\,\ref{sect:pop_syn:results:triple_orb}). Assuming the HJ initially orbited its host star at a separation between 1 and 4 AU, for `2+2' systems the semimajor axis $a_2$ of the stellar binary orbit should satisfy $a_2\lesssim 10^2\,\mathrm{AU}$. Also, the semimajor axis of the `outer' orbit, $a_3$, should satisfy $20\,\mathrm{AU} \lesssim a_3 \lesssim 10^3\,\mathrm{AU}$. 

These constraints are approximately consistent with the three HJs observed so far in hierarchical triple systems, which are all of the `2+2' type. Moreover, a future detection of a HJ violating these conditions would be strong indication for another formation mechanism of HJs in these triples, such as disk migration.

For HJ-hosting `3+1' configurations, the semimajor axis $a_2$ of the first orbit around the star+planet subsystem should satisfy $20\,\mathrm{AU} \lesssim a_2 \lesssim 10^3\,\mathrm{AU}$. For the the second orbit, the semimajor axis $a_3$ should be $10^3 \, \mathrm{AU} \lesssim a_3 \lesssim 10^5 \, \mathrm{AU}$. 

\medskip \noindent 4. For similar ranges of semimajor axes (for details, see \F\,\ref{fig:arranged_sma_distributions_combined_MC03long.eps}), we also expect  planets to be tidally disrupted by the star. This suggests possible enrichment of stars by planets in hierarchical triple systems satisfying specific orbital configurations. Alternatively, if the planet is not completely destroyed during the disruption event it may be partially or completely stripped of its envelope \citep{2013ApJ...762...37L}, producing hot Neptunes or (short-period) super-Earths in these triple systems.

\section*{Acknowledgements}
We thank Dong Lai for stimulating discussions and comments on the manuscript, and the anonymous referee for a helpful report. ASH gratefully acknowledges support from the Institute for Advanced Study.

\bibliographystyle{mnras}
\bibliography{literature}

\label{lastpage}
\end{document}